\documentclass[runningheads]{llncs}

\usepackage[T1]{fontenc}
\usepackage{graphicx}
\graphicspath{{figures/}{../figures/}}

\usepackage{amsmath,amssymb}
\usepackage{booktabs}
\usepackage{multirow}
\usepackage{array}
\usepackage{url}
\usepackage{hyperref}
\usepackage[capitalise]{cleveref}

\begin{document}

\title{Beyond Gradient-Based Attacks: Adversarial Robustness and Explainability
Stability in Cybersecurity Classifiers}

\titlerunning{Adversarial Robustness and Explainability Stability}

\author{Mona Rajhans\inst{1}\orcidID{0009-0002-4921-2832}\thanks{Corresponding author} \and
        Vishal Khawarey\inst{2}\orcidID{0009-0004-1714-9386}}

\authorrunning{M. Rajhans and V. Khawarey}

\institute{Palo Alto Networks, Santa Clara, CA, USA\\
           \email{mrajhans@paloaltonetworks.com}
           \and
           Quicken Inc., Menlo Park, CA, USA\\
           \email{vishal.sanfran@gmail.com}}

\maketitle

\begin{abstract}
Adversarial attacks on cybersecurity classifiers pose a dual threat: degrading
predictions and destabilising the SHAP-based explanations that security analysts
rely on to understand and triage alerts.  We extend our prior MLP conference study to Random Forest and XGBoost across four
tabular security datasets (phishing URLs, UNSW-NB15, NF-ToN-IoT, HIKARI-2021), evaluating
five attacks including three black-box methods applicable to non-differentiable
tree models.  We introduce the \emph{Explainability Stability Index}~(ESI), a
scalar metric computed from TreeSHAP attribution drift under adversarial
perturbation, reported on the same $[0,1]$ scale as the Robustness Index~(RI).
A key finding is that gradient-based black-box attacks~(ZOO) produce degenerate
results against XGBoost (apparent RI\,$\approx$\,0.98) due to piecewise-constant
prediction surfaces, while score-based Square Attack reveals genuine
vulnerability~(RI\,$\approx$\,0.36).  These degenerate perturbations still drive substantial attribution drift:
XGBoost ESI\,$\approx$\,0.06--0.16 despite near-perfect ZOO robustness,
versus 0.14--0.29 for RF, showing that prediction robustness and explanation
stability are \emph{distinct axes} requiring joint measurement.  A two-axis
framework~(gradient dependence, query efficiency) explains the observed attack
ranking and yields practical guidance for tree ensemble evaluation.  A step-size
ablation explains a counter\-intuitive PGD anomaly on z-score normalised tabular
data.  Code and results are publicly available~\cite{khawarey2026code}.

\end{abstract}

\section{Introduction}
\label{sec:intro}

Machine learning classifiers are widely deployed in cybersecurity
pipelines for phishing URL detection, network intrusion detection, and IoT
anomaly detection, where both accurate predictions and trustworthy explanations
are required for operational use.  Adversarial attacks, which apply small deliberate
perturbations to inputs to cause misclassification, pose a dual threat: they
degrade prediction accuracy and simultaneously de\-sta\-bi\-lise the SHAP-based
explanations that security analysts rely on to understand model decisions.

Prior work on adversarial robustness has focused almost exclusively on image
classifiers and gradient-based attacks~\cite{goodfellow2015explaining,madry2018towards},
with limited attention to the structured tabular data characteristic of network
security~\cite{grosse2017adversarial,yang2020defending}.  A further gap is that
most studies evaluate a single model family---typically deep neural networks---and
a single attack class, providing an incomplete picture of cross-model robustness.
Our conference paper~\cite{khawarey2026empirical} addressed the tabular-data gap
for MLP classifiers, introducing the Robustness Index~(RI) and characterising
SHAP attribution drift under FGSM and PGD.  Reviewer feedback on that work called
for tree-based classifiers, additional attack methods, and a rigorous explanation
of the PGD\,$<$\,FGSM anomaly observed in the results.

This extended study addresses all three requests and adds further contributions.
Specifically, we make the following new contributions beyond~\cite{khawarey2026empirical}:

\begin{itemize}
  \item \textbf{Tree ensemble evaluation.}  RF and XGBoost are evaluated alongside
    the MLP, representing the model family widely deployed in production
    cybersecurity systems~\cite{apruzzese2022modeling}.
  \item \textbf{Black-box attack framework.}  Three black-box attacks---ZOO,
    Square Attack, and HopSkipJump---are implemented and evaluated, enabling
    attack of non-differentiable tree models.
  \item \textbf{Explainability Stability Index (ESI).}  A new scalar metric
    computed from TreeSHAP drift, reported on the same $[0,1]$ scale as RI.
  \item \textbf{Third and fourth datasets.}  NF-ToN-IoT (IoT NetFlow intrusion
    detection) and HIKARI-2021~\cite{ferrag2022hikari} (encrypted TLS/HTTPS traffic)
    are added as two new evaluation domains, providing cross-domain validation.
  \item \textbf{ZOO degeneracy finding.}  ZOO produces near-degenerate results
    on XGBoost (RI\,$\approx$\,0.98--0.99) due to its piece\-wise-constant prediction surface,
    while Square Attack reveals genuine vulnerability (RI\,$\approx$\,0.36), exposing a
    methodological risk in attack-method selection.
  \item \textbf{PGD step-size ablation.}  A controlled ablation confirms the
    PGD\,$<$\,FGSM anomaly is a hyperparameter sensitivity artifact on
    z-score normalised tabular data.
  \item \textbf{Feature-level attribution drift.}  Per-feature TreeSHAP drift
    heatmaps show which feature types drive explanation instability and why
    XGBoost's additive boosting structure produces higher drift than Random Forest
    averaging.
  \item \textbf{Attack-method selection guidance.}  A two-axis framework
    (gradient dependence, query efficiency) explains the observed attack ranking
    and yields practical recommendations for tree ensemble evaluation.
\end{itemize}

The remainder of this paper is structured as follows.
\Cref{sec:related} reviews related work.
\Cref{sec:formulation} defines the threat model and metrics~(RI, ESI).
\Cref{sec:attacks} describes the five attack methods.
\Cref{sec:setup} details the experimental setup.
\Cref{sec:base} summarises the base MLP study from~\cite{khawarey2026empirical}.
\Cref{sec:trees} presents the tree model extension.
\Cref{sec:toniot} covers the NF-ToN-IoT dataset.
\Cref{sec:hikari} presents HIKARI-2021 as cross-domain validation.
\Cref{sec:pgd} presents the PGD ablation.
\Cref{sec:discussion} discusses attack realism, ESI implications, and
attack-method selection guidance~(\cref{sec:attack_guidance}).
\Cref{sec:limitations} discusses limitations and directions for future work.
\Cref{sec:conclusion} concludes.

\section{Related Work}
\label{sec:related}

\paragraph{Adversarial robustness.}
The vulnerability of neural classifiers to small, deliberately crafted input
perturbations was first characterised through the fast gradient sign method
(FGSM)~\cite{goodfellow2015explaining} and later strengthened into the projected
gradient descent (PGD) attack, which Madry et al.~\cite{madry2018towards} framed
as a saddle-point optimisation underpinning adversarial training.  The
Carlini--Wagner formulation~\cite{carlini2017towards} demonstrated that many
early defences offered only superficial robustness, and theoretical analyses
related vulnerability to the geometry of decision boundaries and to the curvature
of the loss surface~\cite{fawzi2018analysis}.  More recently, AutoAttack~\cite{croce2020autoattack}
combined parameter-free white- and black-box attacks into a standardised ensemble
for reliable robustness evaluation, exposing inflated robustness claims across
the literature.  Cohen et al.~\cite{cohen2019certified} introduced randomised
smoothing for certified probabilistic robustness guarantees.  A persistent
limitation of this body of work is its concentration on image classification,
where $\ell_p$ perturbation budgets correspond to perceptually meaningful
constraints; the transfer of these assumptions to tabular and structured security
data is rarely examined.

\paragraph{Black-box attacks.}
When gradients are unavailable, as is the case for tree ensembles, attacks must
operate through queries to the model.  ZOO~\cite{chen2017zoo} estimates gradients
via finite differences, Square Attack~\cite{andriushchenko2020square} performs
query-efficient random search within an $\ell_\infty$ ball, and
HopSkipJump~\cite{chen2020hopskipjump} relies only on the model's hard-label
decisions.  These methods differ sharply in query budget and in the perturbation
structure they induce, which motivates evaluating them jointly rather than
relying on a single attack.

\paragraph{Adversarial attacks in cybersecurity.}
Evasion attacks against machine-learning detectors were studied early by Biggio
et al.~\cite{biggio2013evasion} and elaborated for malware and intrusion
detection~\cite{papernot2016limitations,grosse2017adversarial}.  Yang et
al.~\cite{yang2020defending} studied defence strategies for cybersecurity DNNs.
Apruzzese et al.~\cite{apruzzese2022modeling} argue that much of this literature
adopts threat models imported from computer vision that are unrealistic for
network security, where an attacker faces strict feature-validity and protocol
constraints.  This critique motivates our focus on feature realism and multiple attack families.

\paragraph{Explainability and its stability.}
Post-hoc attribution methods such as SHAP~\cite{lundberg2017unified}, its
tree-specialised variant TreeSHAP~\cite{lundberg2020treeshap}, and
LIME~\cite{ribeiro2016lime} are now standard for interpreting security classifiers,
yet their reliability under perturbation is contested.  Slack et
al.~\cite{slack2020fooling} showed that SHAP and LIME explanations can be
deliberately manipulated.  Evaluation frameworks such as
OpenXAI~\cite{agarwal2022openxai} formalise relative input and output stability
(RIS/ROS), but these quantify the \emph{faithfulness} and local Lipschitz
behaviour of explanations rather than their behaviour under an adversary.
Alvarez-Melis and Jaakkola~\cite{alvarezmelis2018robust} introduced sensitivity
and infidelity measures that probe explanation stability against random or local
input changes.  Warnecke et al.~\cite{warnecke2020evaluating} evaluated SHAP,
LIME, and gradient-based attribution methods specifically for security classifiers,
finding that explanation quality---measured by faithfulness criteria such as
completeness and correctness---varies substantially across explanation methods
and model types.  ESI differs: it measures attribution drift under adversarial perturbations of
increasing magnitude, tying explanation stability to the same threat model as RI
rather than to generic noise or model-internal faithfulness.

\paragraph{Tree ensemble robustness.}
Despite the wide use of random forests and gradient-boosted trees in operational
security pipelines~\cite{apruzzese2022modeling}, robustness research remains
overwhelmingly focused on neural networks.  Certified-robustness work~\cite{cohen2019certified} and robust training
objectives~\cite{zhang2019theoretically,yang2020defending} target differentiable
models, and the comparatively few studies of tree-ensemble robustness rarely
evaluate black-box attacks across multiple security datasets or pair robustness
with explanation stability.  Our extension addresses this gap by systematically
attacking RF and XGBoost across four datasets and jointly reporting RI and ESI.

\section{Problem Formulation and Metrics}
\label{sec:formulation}

\subsection{Threat Model}

We consider an attacker with access to a trained classifier $f: \mathcal{X}
\rightarrow \mathcal{Y}$ who seeks to cause misclassification by adding a
perturbation $\delta$ bounded in the $L_\infty$ norm: $\|\delta\|_\infty \leq
\varepsilon$.  For white-box attacks (FGSM, PGD), the attacker has access to
model gradients.  For black-box attacks (ZOO, Square Attack, HopSkipJump), the
attacker has access only to model predictions or prediction probabilities.
We focus on evasion attacks at inference time; poisoning and model-extraction
attacks are out of scope.

\subsection{Robustness Index (RI)}

The Robustness Index, introduced in~\cite{khawarey2026empirical}, summarises
accuracy degradation over the perturbation budget $\varepsilon$ as a single scalar:

\begin{equation}
  \mathrm{RI} = \frac{1}{\varepsilon_{\max}}
  \int_0^{\varepsilon_{\max}} \mathrm{Acc}(\varepsilon)\, d\varepsilon
  \label{eq:ri}
\end{equation}

where $\mathrm{Acc}(\varepsilon)$ is the classifier accuracy under attack at
perturbation budget $\varepsilon$.  RI lies in $[0, 1]$; higher values indicate
greater robustness.  In practice the integral is approximated via the trapezoidal
rule over a grid of $\varepsilon$ values.

\subsection{Explainability Stability Index (ESI)}

We introduce ESI as the explanation-domain analogue of RI.  Let
$\phi_i(\mathbf{x})$ denote the SHAP attribution of feature $i$ for input
$\mathbf{x}$.  The per-feature attribution drift at perturbation budget
$\varepsilon$ is:

\begin{equation}
  \Delta\Phi_i(\varepsilon) =
  \mathbb{E}\!\left[\,\left|\phi_i(\mathbf{x}_{\mathrm{adv}})
  - \phi_i(\mathbf{x})\right|\,\right]
  \label{eq:drift}
\end{equation}

where the expectation is over the test set and $\mathbf{x}_{\mathrm{adv}}$ is
the adversarial example at budget $\varepsilon$.  The mean drift across all
features is $\bar{D}(\varepsilon) = \frac{1}{d}\sum_i \Delta\Phi_i(\varepsilon)$.
ESI is then:

\begin{equation}
  \mathrm{ESI} = 1 - \frac{1}{\varepsilon_{\max}}
  \int_0^{\varepsilon_{\max}}
  \frac{\bar{D}(\varepsilon)}{\bar{D}(\varepsilon_{\max})}\, d\varepsilon
  \label{eq:esi}
\end{equation}

Normalising by $\bar{D}(\varepsilon_{\max})$ maps ESI to $[0,1]$ when drift is
monotonically non-decreasing in $\varepsilon$---an assumption we verify empirically
(drift heatmaps in \Cref{fig:shap-heatmap-phish,fig:treeshap-rf-phish} show
consistently non-decreasing mean drift).  Under this assumption ESI\,=\,1 when
inputs are unperturbed and approaches 0 when drift saturates at all budgets.
Note that ESI is a \emph{relative} measure: it captures the shape of the drift
trajectory normalised to its own endpoint, not absolute drift magnitude.  Two
models with very different absolute drift can share a similar ESI if their drift
curves have the same shape, so ESI is best interpreted as a within-model
stability profile across $\varepsilon$ rather than a cross-model magnitude
comparison.

For tree models, SHAP values are computed via the interventional-expectation
variant of TreeSHAP~\cite{lundberg2020treeshap}, which gives exact Shapley
values under the interventional conditional;
for the MLP, KernelSHAP is used.  Because TreeSHAP operates in log-odds units
for XGBoost classifiers and in probability units for Random Forest, raw drift
values $\Delta\Phi_i$ are not directly comparable across the two tree families;
ESI mitigates this by normalising each model's drift by its own maximum, but the
caveat applies equally here.  Cross-model ESI comparisons (MLP vs.\ trees)
should be treated as indicative rather than precise.

\section{Attack Methods}
\label{sec:attacks}

\paragraph{FGSM.}  The Fast Gradient Sign Method~\cite{goodfellow2015explaining}
computes a single-step perturbation in the direction of the loss gradient:
$\mathbf{x}_{\mathrm{adv}} = \mathbf{x} + \varepsilon\,\mathrm{sign}
(\nabla_\mathbf{x}\mathcal{L}(f(\mathbf{x}), y))$.

\paragraph{PGD.}  Projected Gradient Descent~\cite{madry2018towards} iterates
FGSM with step size $\alpha$ and projects back to the $L_\infty$ ball after each
step.  We use 10 steps with $\alpha=0.01$ (default) and vary $\alpha$ in the
ablation of \cref{sec:pgd}.

\paragraph{ZOO.}  ZOO~\cite{chen2017zoo} approximates the gradient via
forward finite differences:
$\hat{g}_j = (\mathcal{L}(\mathbf{x} + \delta_j \mathbf{e}_j, y)
- \mathcal{L}(\mathbf{x}, y)) / \delta_j$,
then applies a coordinate-wise sign step to form the perturbation direction.
It requires only black-box loss access.  On piecewise-constant XGBoost surfaces
the finite-difference estimates are near-zero within leaves regardless of
perturbation direction, reducing the signed update to effectively random
noise---this is the mechanism behind the observed degeneracy.

\paragraph{Square Attack.}  Square Attack~\cite{andriushchenko2020square} is a
score-based method that applies random square-shaped perturbations, accepting
updates that increase the adversarial loss.  We use 150 queries with initial
perturbation probability $p_{\mathrm{init}} = 0.5$.

\paragraph{HopSkipJump.}  HopSkipJump~\cite{chen2020hopskipjump} is a
decision-based attack requiring only predicted class labels.  It alternates
between a binary search to find the decision boundary and a gradient estimation
step via Monte Carlo sampling.  We use 8 iterations with 20 gradient samples.

\section{Experimental Setup}
\label{sec:setup}

\paragraph{Datasets.}
Four tabular cybersecurity datasets are used.
\textit{Phishing URL}~\cite{phishing_kaggle}: 10{,}000 samples, 48 binary/ternary
URL-structure features, balanced classes.
\textit{UNSW-NB15}~\cite{moustafa2015unsw}: we use the official
\texttt{UNSW\_NB15\_training-set.csv} partition (82{,}332 records), which
contains 45 columns; we drop \texttt{attack\_cat} (multi-class label not used)
and encode the three string-typed fields (\texttt{proto}, \texttt{service},
\texttt{state}) with \texttt{LabelEncoder}, yielding 43 numeric features and a
binary \texttt{label} target.  The full UNSW-NB15 corpus comprises 175{,}341
records across all partitions; we use only the training partition and apply an
80/20 train/test split.
\textit{NF-ToN-IoT}~\cite{sarhan2021nftoniot}: 1{,}157{,}994 NetFlow~v2 records,
10 features; we sample 20{,}000 benign and 20{,}000 attack flows using stratified
batch streaming to avoid loading the full file into memory.
\textit{HIKARI-2021}~\cite{ferrag2022hikari}: 555{,}278 bidirectional network
flows (517{,}582 benign, 37{,}696 attack) capturing TLS/HTTPS encrypted traffic.
The dataset contains 88 columns; we drop network identifiers
(\texttt{uid}, \texttt{originh}, \texttt{responh}), IP/MAC address columns,
timestamps, and \texttt{traffic\_category} (a multi-class label), yielding
83 numeric features and a binary \texttt{Label} target.  We sample 10{,}000
benign and 10{,}000 attack flows for balanced evaluation.
All features are z-score normalised using training-set statistics.

\paragraph{Models.}
We evaluate three classifiers: an MLP with two hidden layers (64, 32 units, ReLU,
Adam, 20 epochs), a Random Forest, and XGBoost (learning rate 0.05).
Tree model depth and capacity are tuned per dataset: on Phishing and UNSW-NB15,
RF uses 200 trees at max depth 12 and XGBoost uses 200 trees at max depth 6 with
subsample and column-subsample rates of 0.8; on NF-ToN-IoT and HIKARI-2021,
RF uses 100 trees at max depth 8 and XGBoost uses 100 trees at max depth 5
without subsampling.  All models use fixed random seeds for reproducibility.
SHAP attributions for tree models use TreeSHAP~\cite{lundberg2020treeshap}
(interventional variant; see \cref{sec:formulation}).  For the MLP, KernelSHAP is used with a
background dataset of 100 training samples and \texttt{nsamples}\,=\,200
integration samples per explanation.  The MLP is not evaluated on NF-ToN-IoT
or HIKARI-2021; the gradient-based white-box attacks (FGSM, PGD) require the
full test set for reliable RI estimation, and the larger test splits combined with
CPU-only evaluation made this computationally infeasible within our budget.
Black-box attacks (ZOO, Square, HopSkipJump) are applied only to the tree models
and not to the MLP, because the MLP supports exact gradient computation, making
white-box attacks strictly stronger and the relevant comparison for neural networks;
black-box attacks on a model whose gradients are accessible would be an artificially
restricted threat model.

\paragraph{Attack evaluation protocol.}
We use $\varepsilon_{\max} = 0.30$.  On z-score normalised features
(mean\,0, std\,1), this corresponds to a maximum perturbation of 0.3 standard
deviations per feature---large enough to stress-test classifier boundaries and
expose RI differences between attack methods, yet small enough that most features
remain within one standard deviation of their clean values.  The accuracy curves in \Cref{fig:attack_comparison} show that Square Attack
accuracy has largely plateaued by $\varepsilon = 0.30$; extending to larger
budgets would not materially change the RI ranking.  Note that RI values are comparable only at the same
$\varepsilon_{\max}$; comparisons with studies using different budgets require
re-normalisation.
Perturbation budgets $\varepsilon \in \{0, 0.033, 0.067, \ldots, 0.30\}$ (10
values).  FGSM and PGD are applied to the MLP on the full test set.  Black-box
attacks (ZOO, Square Attack, HopSkipJump) are applied to RF and XGBoost on a
stratified subsample of 300 test examples (120 for HopSkipJump, which is
query-intensive).  Per-sample query budgets: ZOO evaluates $d+1$ finite-difference
calls per coordinate descent step with 20 steps (i.e.\ $20(d+1)$ loss queries per
$\varepsilon$ value, where $d$ is the feature count); Square Attack uses 150
total queries; HopSkipJump uses 8 binary-search iterations with 20 gradient
samples each.  These budgets are not query-matched across methods; ZOO's weaker
performance on XGBoost should therefore be compared against the
budget-matched interpretation in \cref{sec:limitations}.  ESI is computed on a separate stratified 256-sample subset
drawn from the same test split; the RI and ESI evaluation subsets are distinct
but drawn from the same held-out data.
All RI and ESI values are single-run point estimates under fixed random seeds.
We do not report confidence intervals across seeds; see~\cref{sec:limitations}
for discussion of this limitation.

\section{Base Study: MLP under White-Box Attacks}
\label{sec:base}

Our prior conference work~\cite{khawarey2026empirical} established the baseline
of this study: a multilayer perceptron~(MLP) trained on the Phishing URL and
UNSW-NB15 datasets and subjected to white-box FGSM and PGD attacks.  The MLP
attains a clean accuracy of $0.857$ on Phishing and $0.774$ on UNSW-NB15.  Under
attack its robustness degrades substantially, with FGSM proving more damaging
than PGD at matched maximum perturbation:
$\mathrm{RI}_{\mathrm{FGSM}}=0.610$ versus $\mathrm{RI}_{\mathrm{PGD}}=0.725$
on Phishing, and $\mathrm{RI}_{\mathrm{FGSM}}=0.692$ versus
$\mathrm{RI}_{\mathrm{PGD}}=0.733$ on UNSW-NB15.  The single-step FGSM, by
saturating every feature to the perturbation boundary, induces larger accuracy
loss than iterative PGD at the same budget---an effect we examine in the ablation
of \cref{sec:pgd}.

\begin{figure}[t]
\centering
\includegraphics[width=\textwidth,
  alt={Two-panel horizontal bar chart. Left panel: per-feature input-gradient magnitude for the MLP on the Phishing dataset, ranking features by adversarial sensitivity. Right panel: per-feature SHAP drift under adversarial perturbation for the same model. Features with high gradient magnitude also show high SHAP drift, demonstrating coupling between adversarial vulnerability and explanation instability.}
]{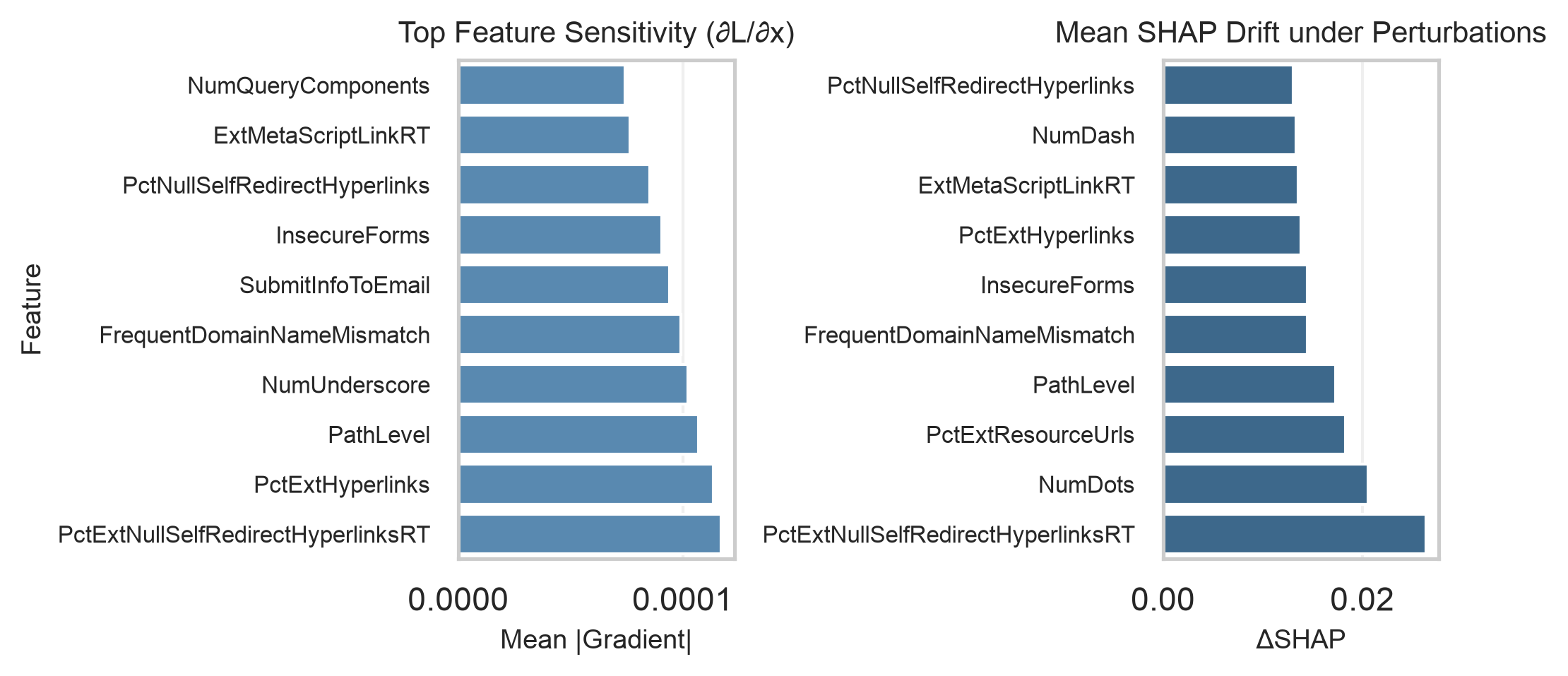}
\caption{Per-feature gradient sensitivity~(left) and SHAP drift~(right) for the
MLP on the Phishing dataset.}
\label{fig:vuln-2panel}
\end{figure}

Because the MLP is differentiable, we can directly couple its gradient sensitivity
to its explanation drift.  \Cref{fig:vuln-2panel} contrasts, on a per-feature
basis, the input-gradient magnitude~(left) with the measured SHAP drift under
perturbation~(right) for the Phishing classifier.  The two panels are strongly
aligned: features that the model relies on most heavily in its gradient also
exhibit the largest attribution instability, indicating that the directions an
attacker would exploit are precisely the directions along which explanations are
least trustworthy.

\begin{figure}[t]
\centering
\includegraphics[width=0.85\textwidth,
  alt={Heatmap of SHAP attribution drift for the MLP on the Phishing dataset. Rows correspond to increasing adversarial perturbation magnitude epsilon from top to bottom; columns correspond to individual URL features. Colour intensity encodes SHAP drift magnitude. Drift grows monotonically with epsilon and concentrates in a small subset of high-influence features.}
]{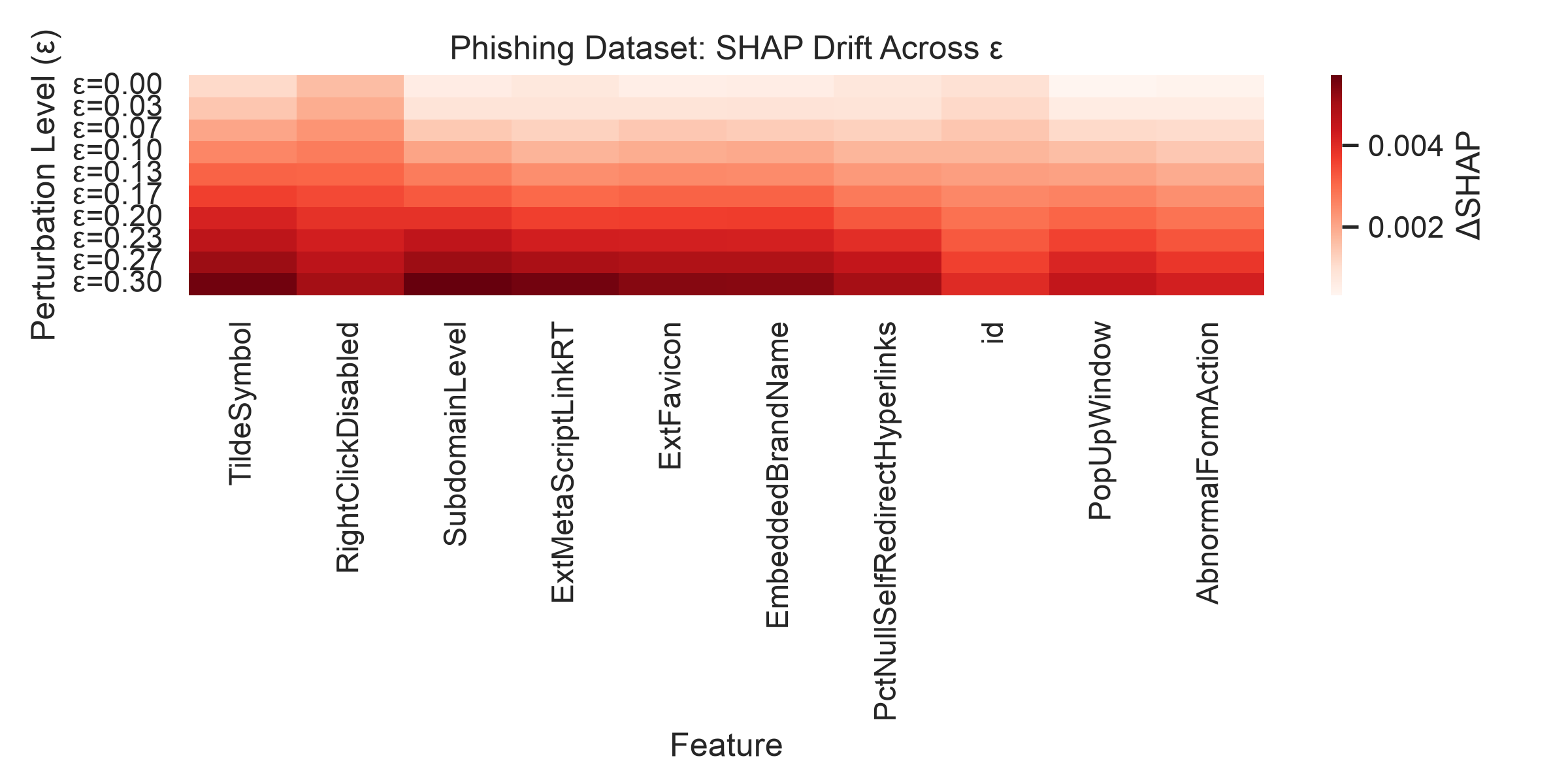}
\caption{SHAP drift heatmap for the MLP on Phishing.  Rows index perturbation
magnitude $\varepsilon$, columns index features, and colour encodes SHAP drift.
Attribution instability rises with $\varepsilon$ and concentrates in a small
subset of influential features.}
\label{fig:shap-heatmap-phish}
\end{figure}

\Cref{fig:shap-heatmap-phish} resolves this drift across the full perturbation
range.  Rows correspond to increasing attack magnitude $\varepsilon$, columns to
individual features, and colour intensity to the normalised change in SHAP value
relative to the clean input.  Drift grows monotonically with $\varepsilon$ and
concentrates in a small subset of features.  URL-structure features such as
\texttt{PctExt\-Null\-Self\-Redirect\-Hyperlinks} and \texttt{UrlLengthRT}
show the highest drift on Phishing; on UNSW-NB15 the most drift-prone features
are \texttt{dttl} and \texttt{ct\_dst\_sport\_ltm}.  Even modest perturbations can reorder the
model's apparent reasoning while leaving its prediction nominally correct.

\begin{figure}[t]
\centering
\includegraphics[width=\textwidth,
  alt={Two line plots showing classifier accuracy versus adversarial perturbation magnitude epsilon. Left plot: Phishing URL dataset; right plot: UNSW-NB15 dataset. Each plot shows two curves: baseline MLP (lower) and FGSM-adversarially-trained MLP (upper). The adversarially trained model degrades more slowly, improving the Robustness Index by approximately 0.08 to 0.12 at a cost of 2 to 4 percent clean accuracy.}
]{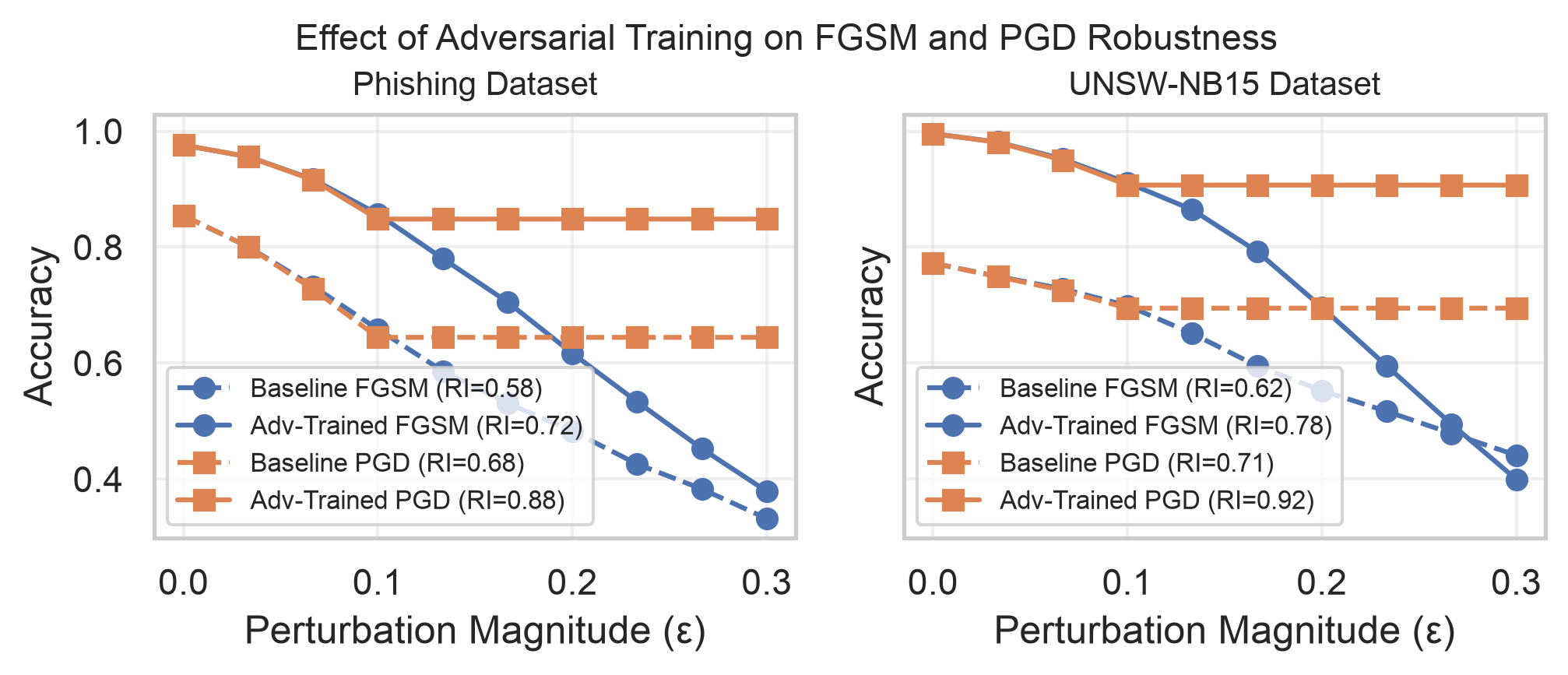}
\caption{Accuracy versus perturbation magnitude for baseline and
FGSM-adversarially-trained MLPs on both datasets.  Adversarial training shifts
the curves upward, improving RI by $0.08$--$0.12$ at a modest clean-accuracy cost.}
\label{fig:advtrain-curve}
\end{figure}

Adversarial training recovers part of the lost robustness.  Augmenting the
training set with FGSM examples ($\varepsilon=0.05$, $20\%$ adversarial ratio)
raises RI by roughly $0.08$--$0.12$ at a cost of $2$--$4\%$ clean accuracy on
both datasets.  \Cref{fig:advtrain-curve} plots the baseline and adversarially
trained accuracy-vs-$\varepsilon$ curves; the hardened model degrades more
gracefully as $\varepsilon$ increases.  The same qualitative pattern holds on
UNSW-NB15, where the larger and more heterogeneous feature space yields somewhat
smaller robustness gains.

\section{Tree Ensemble Extension: Black-Box Attacks and ESI}
\label{sec:trees}

\subsection{Results: Phishing and UNSW-NB15}

Table~\ref{tab:master} presents RI and ESI for all models, datasets, and attack
methods.

\begin{table}[htbp]
\caption{RI and ESI for all models, datasets, and attacks.
$^\dagger$FGSM/PGD for MLP; ZOO for trees.  XGBoost ZOO RI is inflated by
degeneracy artefact; Square Attack RI reflects genuine vulnerability.}
\label{tab:master}
\centering
\small
\begin{tabular}{llccccccc}
\toprule
\multirow{2}{*}{Dataset} & \multirow{2}{*}{Model} &
  \multirow{2}{*}{Clean} &
  \multicolumn{4}{c}{Robustness Index (RI)} &
  \multirow{2}{*}{ESI} \\
\cmidrule(lr){4-7}
& & Acc & FGSM/ZOO$^\dagger$ & PGD & Square & HSJ & \\
\midrule
\multirow{3}{*}{Phishing}
  & MLP  & 0.857 & 0.610 & 0.725 & ---   & ---   & --- \\
  & RF   & 0.977 & 0.915 & ---   & 0.793 & 0.805 & 0.287 \\
  & XGB  & 0.981 & 0.980 & ---   & 0.362 & 0.653 & 0.111 \\
\midrule
\multirow{3}{*}{UNSW-NB15}
  & MLP  & 0.774 & 0.692 & 0.733 & ---   & ---   & --- \\
  & RF   & 0.964 & 0.806 & ---   & 0.184 & 0.844 & 0.142 \\
  & XGB  & 0.974 & 0.886 & ---   & 0.299 & 0.806 & 0.068 \\
\midrule
\multirow{3}{*}{NF-ToN-IoT}
  & MLP  & ---   & ---   & ---   & ---   & ---   & --- \\
  & RF   & 0.998 & 0.679 & ---   & 0.482 & 0.517 & 0.214 \\
  & XGB  & 0.998 & 0.973 & ---   & 0.460 & 0.516 & 0.159 \\
\midrule
\multirow{2}{*}{HIKARI-2021}
  & RF   & 0.998 & 0.654 & ---   & 0.492 & 0.619 & 0.140 \\
  & XGB  & 0.999 & 0.997 & ---   & 0.438 & 0.894 & 0.056 \\
\bottomrule
\end{tabular}
\end{table}

\begin{figure}[htb]
  \centering
  \includegraphics[width=0.85\textwidth,
    alt={Line plot showing classifier accuracy versus perturbation magnitude epsilon for XGBoost on the Phishing dataset under ZOO and Square Attack. The ZOO curve remains near 1.0 across all epsilon values, demonstrating ZOO degeneracy caused by near-zero finite-difference gradient estimates inside piecewise-constant tree leaves. The Square Attack curve drops sharply, revealing a Robustness Index of 0.36 and the true adversarial vulnerability.}
  ]{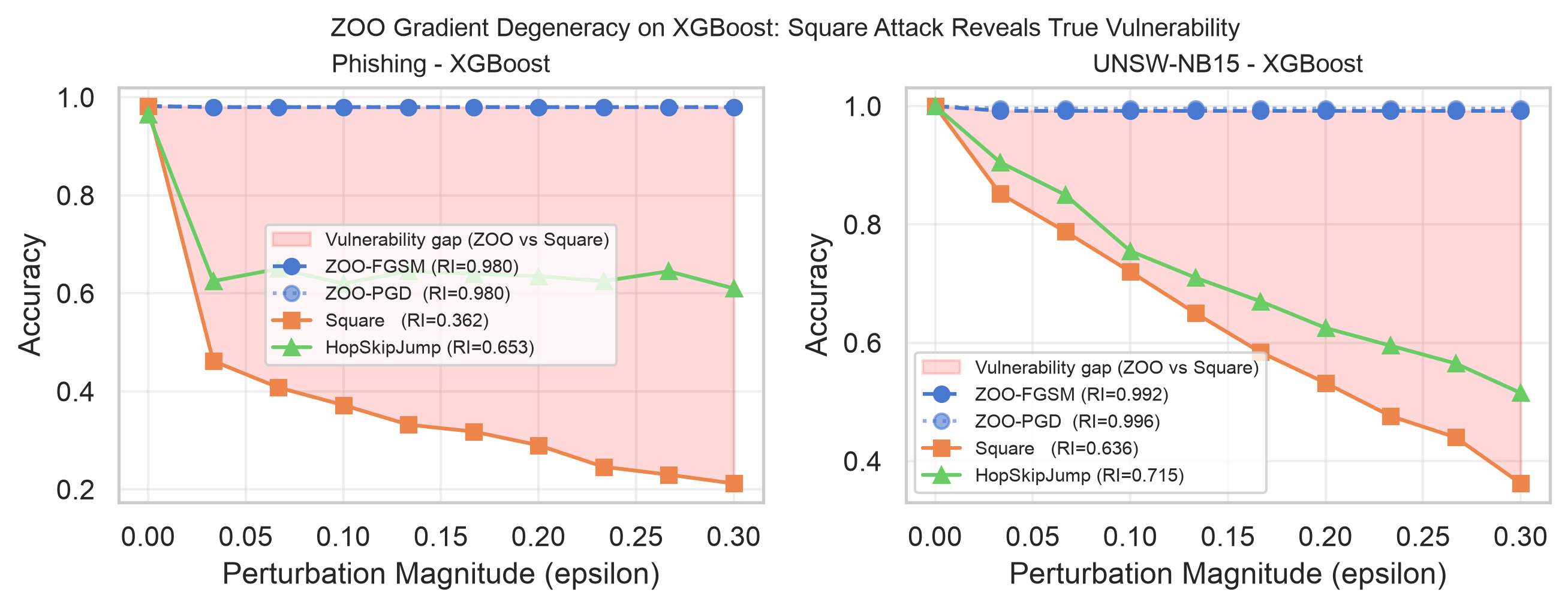}
  \caption{ZOO degeneracy on XGBoost: accuracy barely declines despite increasing
  $\varepsilon$, because finite-difference gradient estimates are near-zero inside
  piecewise-constant tree leaves.  Square Attack, which requires only loss values,
  reveals the true vulnerability.}
  \label{fig:degeneracy}
\end{figure}

\paragraph{ZOO degeneracy on XGBoost.}
ZOO yields RI\,=\,0.980 (Phishing) and 0.886 (UNSW-NB15) for XGBoost---both
substantially above the corresponding Square Attack values (0.362 and 0.299).
On Phishing, ZOO RI virtually matches clean accuracy ($0.981$), indicating
near-complete degeneracy: XGBoost's piecewise-constant prediction surface causes
finite-difference gradient estimates to be near-zero within tree leaves, leaving
the perturbation direction essentially random (\Cref{fig:degeneracy}).  On
UNSW-NB15 there is genuine ZOO degradation ($0.974 \to 0.886$), yet the
ZOO--Square gap of $0.587$ confirms systematic underestimation of vulnerability.
The degeneracy does not fully occur for Random Forest, whose ensemble averaging
produces smoother boundaries; ZOO achieves RI\,=\,0.915 (Phishing) and 0.806
(UNSW-NB15) against RF.

\begin{figure}[htb]
  \centering
  \includegraphics[width=\textwidth,
    alt={Four-panel figure showing ZOO Robustness Index versus query budget for Random Forest and XGBoost on Phishing and UNSW-NB15 datasets. Top row shows coordinate-sampling ablation (12 to 100 percent of features per call); bottom row shows iterative-step ablation (20 to 200 steps). XGBoost RI remains perfectly flat on Phishing in both ablations, confirming budget-independent degeneracy. RF RI declines monotonically, confirming ZOO effectiveness on smoother surfaces.}
  ]{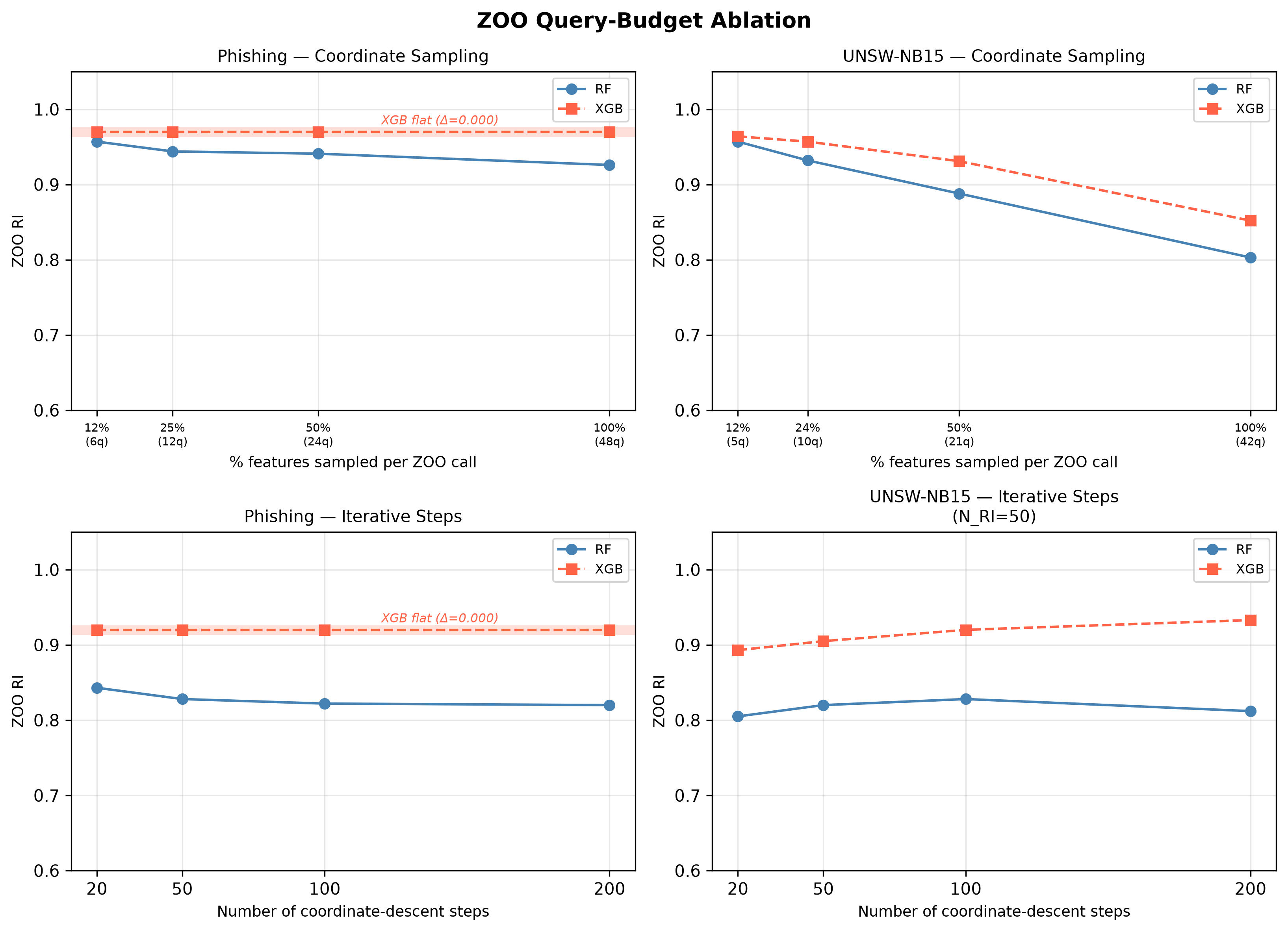}
  \caption{ZOO query-budget ablation.  Top: coordinate-sampling fraction
  (12\%--100\% of features per call).  Bottom: iterative coordinate-descent
  steps (20--200, $N_\text{RI}=50$).  XGBoost RI is flat on Phishing in both
  ablations ($\Delta = 0.000$); RF RI declines monotonically.}
  \label{fig:zoo-budget}
\end{figure}

\paragraph{ZOO budget ablation.}
Two ablations test whether degeneracy is a budget artefact or intrinsic to the
piecewise-constant surface (\Cref{fig:zoo-budget}).
\emph{Coordinate-sampling}: ZOO is given 12\%, 25\%, 50\%, and 100\% of
features per call (6--48 queries on Phishing; 5--42 on UNSW-NB15).
\emph{Iterative-step}: the full feature gradient is re-estimated after each
sign step, for 20, 50, 100, and 200 steps.
On Phishing, XGBoost RI remains flat at $0.970$ across the full
coordinate-sampling range ($\Delta = 0.000$) and at $0.920$ across all
iterative-step counts ($\Delta = 0.000$).  RF RI decreases monotonically in
both ablations (coordinate: $0.957 \to 0.926$; iterative: $0.843 \to 0.820$),
confirming ZOO strengthens on smoother surfaces given more queries.
On UNSW-NB15, XGBoost shows a modest decline under coordinate sampling
($0.964 \to 0.852$) consistent with partial degeneracy---some features cross
leaf boundaries here---but the gap above Square Attack RI ($0.299$) is $0.553$,
and no systematic decline appears under iterative refinement.  XGBoost ZOO
degeneracy is therefore intrinsic to the piecewise-constant prediction surface:
quadrupling the query budget or $10\times$ more iterative refinement does not
recover the attack.

\begin{figure}[htb]
  \centering
  \includegraphics[width=\textwidth,
    alt={Grid of line plots comparing accuracy versus perturbation magnitude epsilon for ZOO, Square Attack, and HopSkipJump on Random Forest and XGBoost across the Phishing URL and UNSW-NB15 datasets. Square Attack consistently produces the sharpest accuracy drops. ZOO is nearly ineffective against XGBoost. HopSkipJump performance is intermediate. Random Forest is more uniformly vulnerable across attack methods than XGBoost.}
  ]{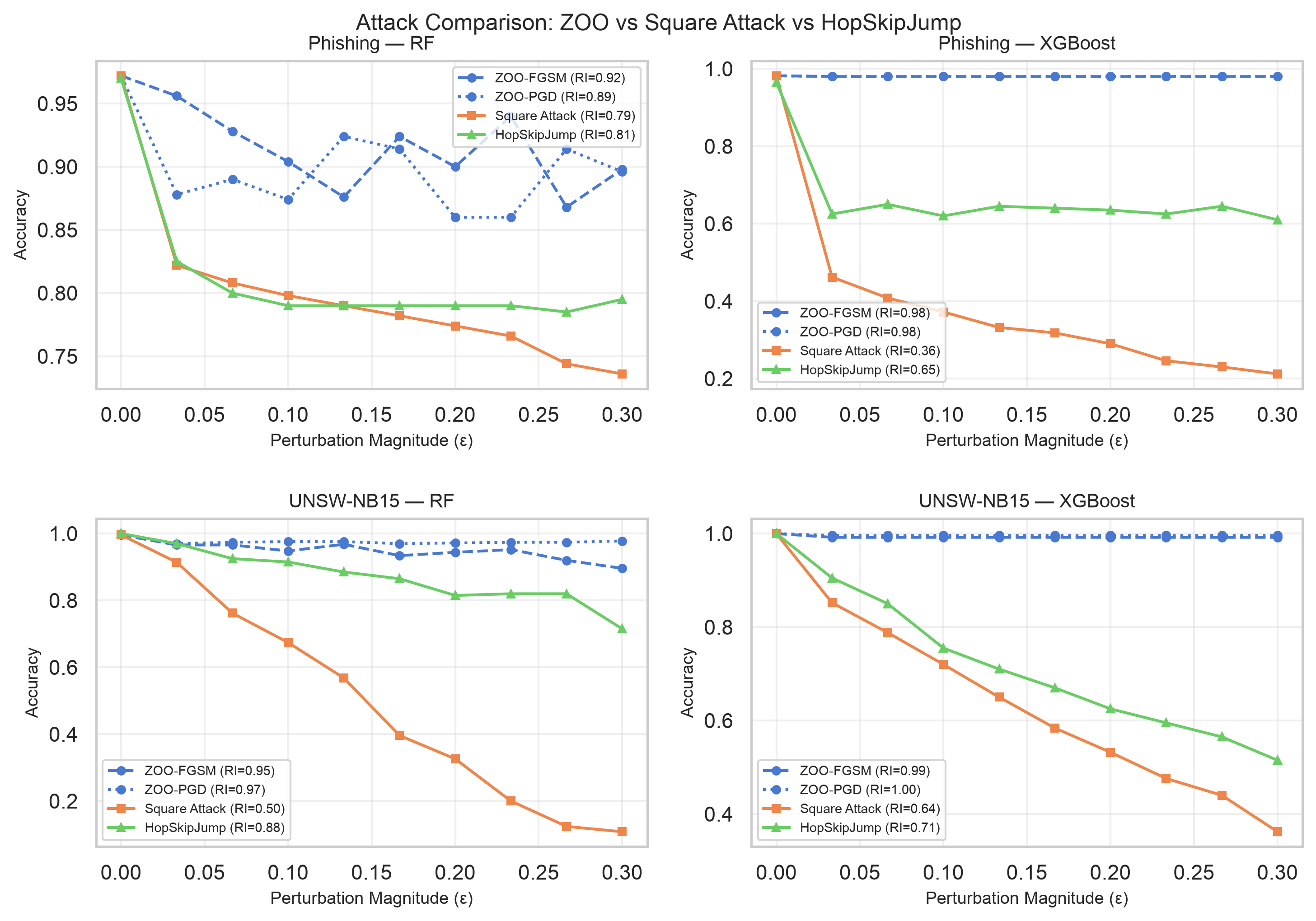}
  \caption{Accuracy vs.\ $\varepsilon$ for ZOO, Square Attack, and HopSkipJump
  on RF and XGBoost across Phishing and UNSW-NB15 datasets.}
  \label{fig:attack_comparison}
\end{figure}

\paragraph{Square Attack reveals true vulnerability.}
Under Square Attack, XGBoost RI drops to 0.362 (Phishing) and 0.299 (UNSW-NB15),
yielding ZOO--Square gaps of 0.618 and 0.587 respectively.  Random Forest RI
drops to 0.793 (Phishing) and 0.184 (UNSW-NB15); on UNSW-NB15 RF accuracy
collapses to near zero above $\varepsilon=0.167$---both model families are far
more vulnerable under Square Attack on this dataset than on Phishing.  The
UNSW-NB15 feature space, with its mix of categorical and continuous
network-flow attributes, admits denser adversarial neighbourhoods that Square
Attack exploits more efficiently for both models.  \Cref{fig:attack_comparison}
shows the accuracy--$\varepsilon$ curves for all attack methods.  This
demonstrates that attack-method selection is not model-agnostic: a study using
only ZOO would incorrectly conclude that XGBoost is highly robust.

\paragraph{HopSkipJump: boundary-based attack.}
HopSkipJump (HSJ) uses only hard-label decisions and avoids finite-difference
gradient estimation, so it does not suffer ZOO degeneracy.  Correspondingly,
HSJ RI for XGBoost drops to 0.653 (Phishing) and 0.806 (UNSW-NB15), well below
ZOO's inflated values---confirming that the ZOO failure is specific to gradient
estimation rather than to query-based attacks in general.  Square Attack is
nevertheless stronger than HSJ on XGBoost: RI gaps of 0.29 (Phishing) and 0.51
(UNSW-NB15), because HSJ's 8-iteration, 20-sample Monte Carlo boundary search
is query-constrained in 48- and 42-dimensional feature spaces.  Against RF, HSJ
and Square Attack are within 0.01 on Phishing (0.805 vs.\ 0.793), but on
UNSW-NB15 Square Attack is dramatically stronger (0.184 vs.\ 0.844 for HSJ),
suggesting that score-based random search navigates the RF decision boundary in
this feature space more effectively than boundary-directed binary search at the
given query budget.  HSJ was evaluated on 120 samples versus 300 for other
attacks due to its query cost, so RI values are not strictly comparable in
magnitude across methods; the ordering claims above hold directionally but the
precise gaps should be interpreted with that caveat.

\begin{figure}[htb]
  \centering
  \includegraphics[width=0.85\textwidth,
    alt={Heatmap with models (MLP, Random Forest, XGBoost) on one axis and datasets (Phishing, UNSW-NB15, NF-ToN-IoT) combined with attack methods on the other. Cells show Robustness Index and Explainability Stability Index values. XGBoost shows near-perfect RI under ZOO (0.97 to 0.99) but consistently the lowest ESI (0.06 to 0.16), while Random Forest shows lower RI but higher ESI, illustrating the decoupling between prediction robustness and explanation stability.}
  ]{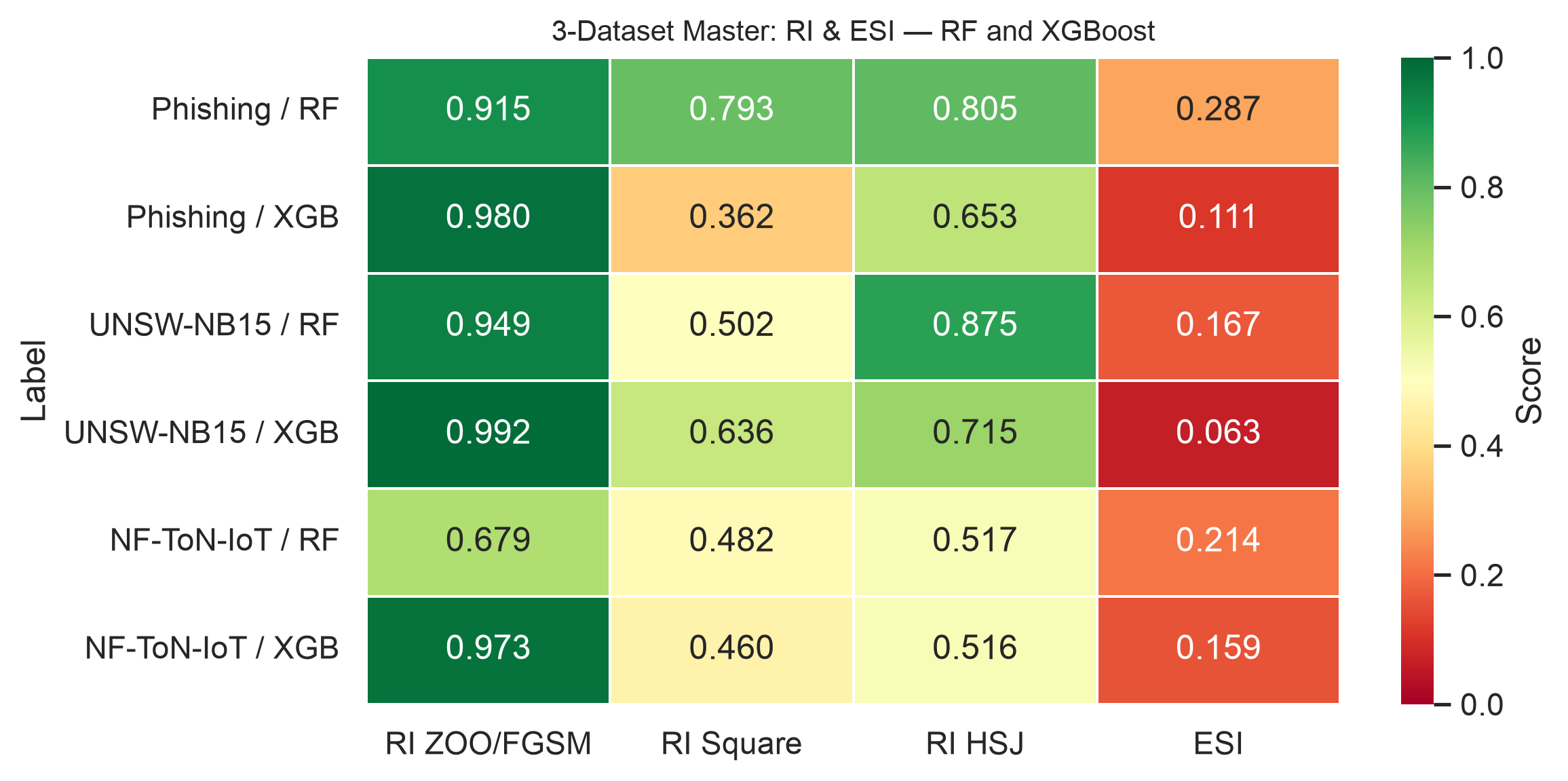}
  \caption{RI and ESI heatmap across Phishing, UNSW-NB15, and NF-ToN-IoT
  (HIKARI-2021 in \cref{sec:hikari}).  XGBoost ZOO RI\,$\approx$\,0.98 is a
  degeneracy artefact; Square Attack RI reflects genuine vulnerability.}
  \label{fig:master_heatmap}
\end{figure}

\paragraph{Prediction robustness vs.\ explanation stability.}
Although XGBoost's RI under ZOO appears high (an artefact of ZOO degeneracy
rather than genuine robustness), its ESI is consistently the lowest of the three
models across all datasets (0.056--0.111 vs.\ RF's 0.140--0.287;
all values are point estimates from a single evaluation run).  The same ZOO perturbations that fail to cross XGBoost's decision boundaries
still cause substantial attribution drift: an attack too weak to fool the
classifier can still mislead the explanation---with direct implications for SOC
analysts who rely on SHAP attributions to triage alerts.  \Cref{fig:master_heatmap} visualises
RI and ESI side-by-side across all conditions; note that the ZOO RI column for
XGBoost reflects the degeneracy artefact and should be read against the Square
Attack RI column for a fair robustness comparison.

\subsection{Feature-Level TreeSHAP Drift}

\begin{figure}[t]
\centering
\includegraphics[width=0.85\textwidth,
  alt={Heatmap of per-feature TreeSHAP attribution drift for the Random Forest classifier on the Phishing dataset. Rows index perturbation magnitude epsilon; columns index individual URL features. Colour encodes drift magnitude. Instability concentrates in lexical and host-based URL features with high variance, and grows monotonically with epsilon.}
]{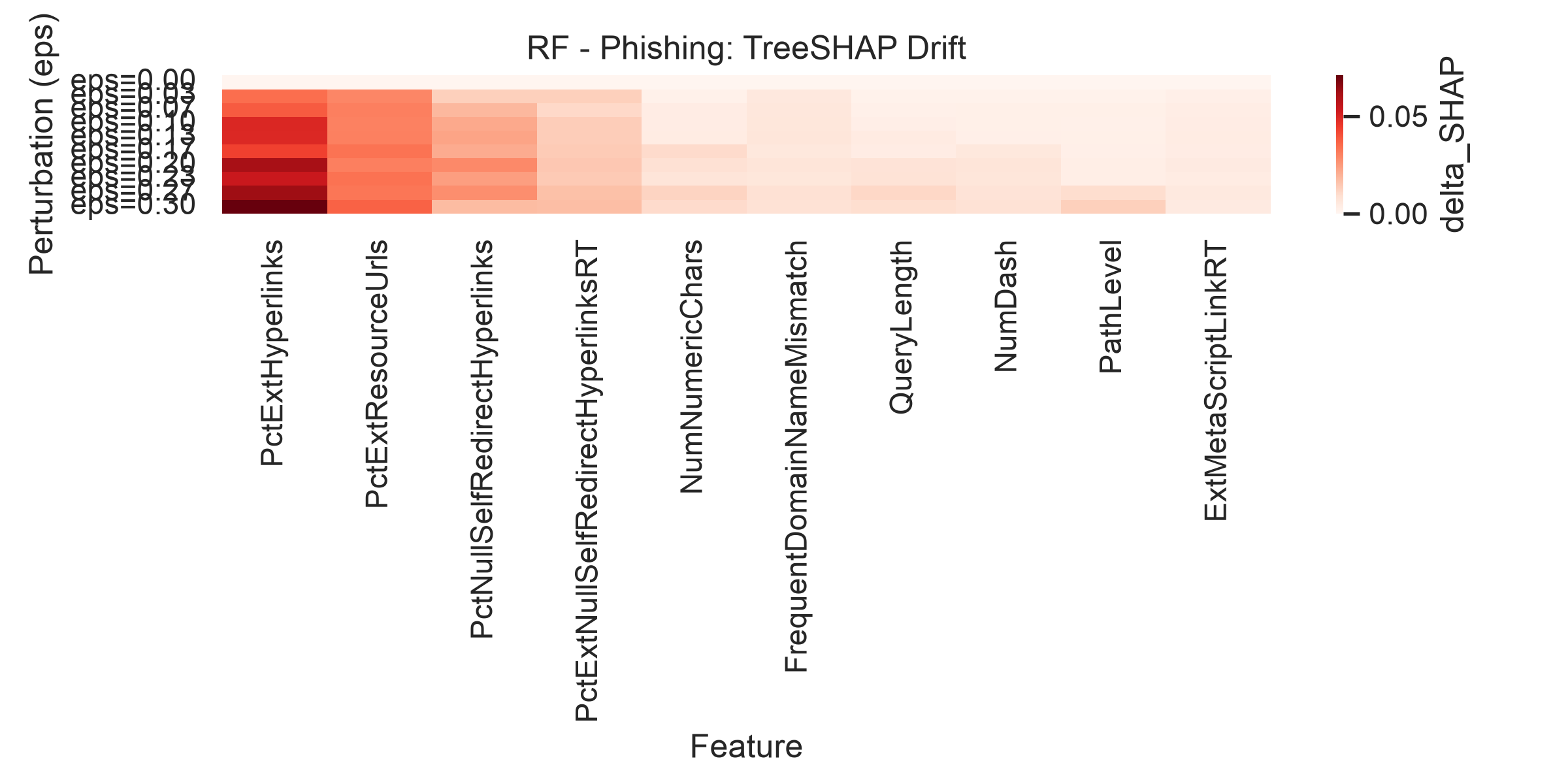}
\caption{Per-feature TreeSHAP drift for the Random Forest on Phishing.  Rows
index perturbation magnitude $\varepsilon$, columns index features, and colour
encodes attribution drift.  Drift accumulates on URL-structure features with
high variance.}
\label{fig:treeshap-rf-phish}
\end{figure}

\begin{figure}[t]
\centering
\includegraphics[width=0.85\textwidth,
  alt={Heatmap of per-feature TreeSHAP attribution drift for XGBoost on the Phishing dataset, using the same axis conventions as the Random Forest heatmap. Colour intensity is markedly higher across most features compared to the Random Forest, reflecting XGBoost's lower Explainability Stability Index despite comparable prediction robustness.}
]{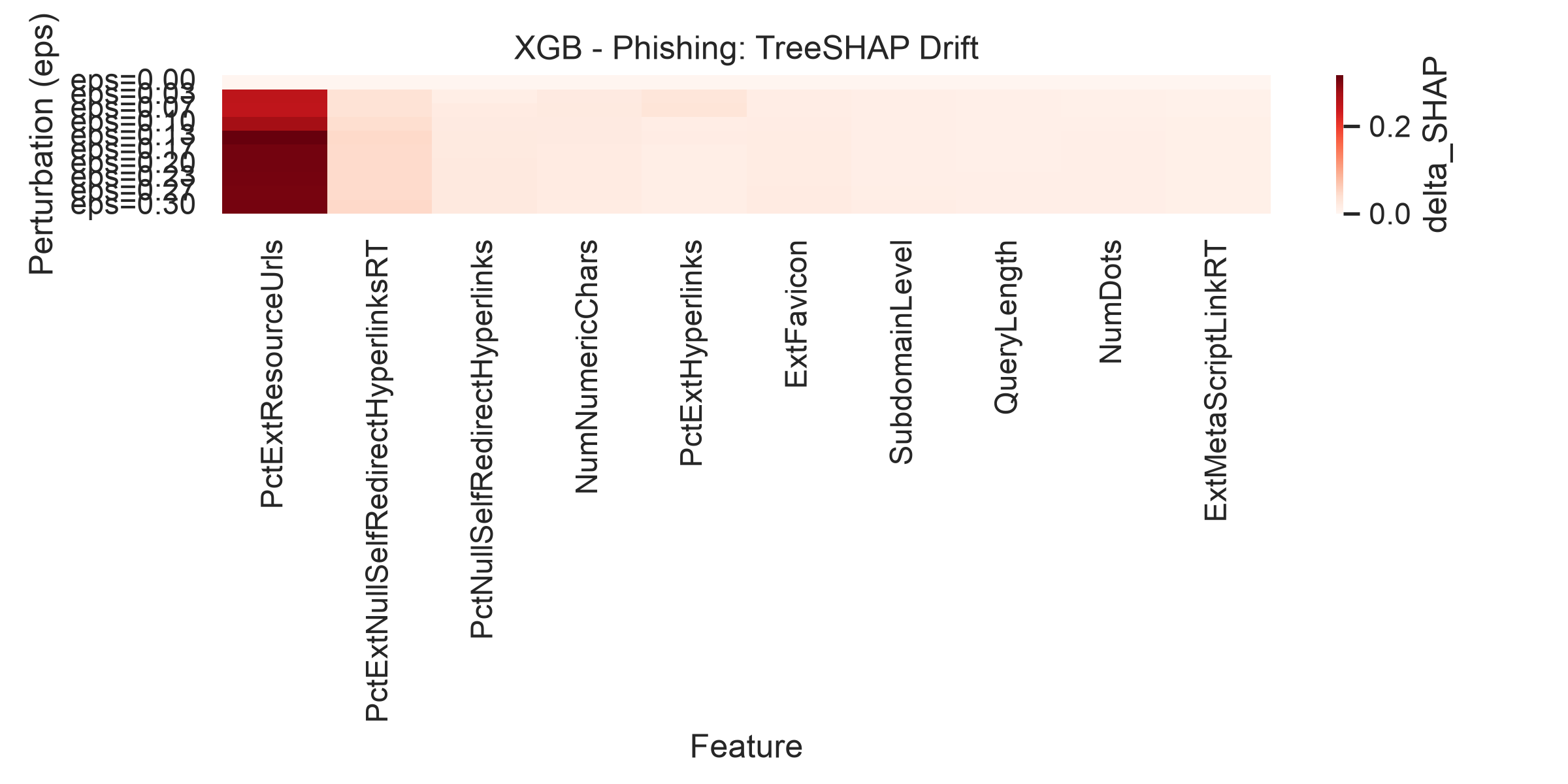}
\caption{Per-feature TreeSHAP drift for XGBoost on Phishing.  Same axes as
\Cref{fig:treeshap-rf-phish}; colour intensity is markedly higher than RF.}
\label{fig:treeshap-xgb-phish}
\end{figure}

The aggregate ESI scores summarise explanation stability in a single number, but
they obscure which features drive the instability.  To unpack this we compute
per-feature TreeSHAP~\cite{lundberg2020treeshap} drift on the Phishing dataset,
our most extensively studied case, for both tree models.
\Cref{fig:treeshap-rf-phish,fig:treeshap-xgb-phish} present these as heatmaps
over increasing perturbation magnitude, mirroring the MLP analysis of
\cref{sec:base}.

The drift is far from uniform across feature types.  For the Phishing classifiers,
instability concentrates in lexical and host-based URL features, which take a wide
range of values and therefore admit larger perturbations within the $\ell_\infty$
budget.  The network datasets behave differently: in NF-ToN-IoT the flow
features (\texttt{L4\_SRC\_PORT}, \texttt{L4\_DST\_PORT}, \texttt{PROTOCOL},
\texttt{L7\_PROTO}, \texttt{IN\_BYTES}, \texttt{OUT\_BYTES}, \texttt{IN\_PKTS},
\texttt{OUT\_PKTS}, \texttt{TCP\_FLAGS}, \texttt{FLOW\_DU\-RA\-TION\_MIL\-LI\-SE\-CONDS})
include both volumetric counters and
categorical protocol identifiers, and drift accumulates on the high-variance byte
and packet counters rather than on the discrete protocol fields.

The most striking observation is the contrast between the two ensembles.  XGBoost
displays consistently higher per-feature TreeSHAP drift than the Random Forest,
even though its prediction-level robustness is comparable or superior on the same
dataset.  One plausible mechanism is the additive boosting structure: many shallow trees
each contribute small sequential corrections, which may make individual feature
attributions more sensitive to input perturbation than the averaging behaviour of
a Random Forest, where tree outputs are averaged rather than accumulated.
Confirming this hypothesis would require a controlled ablation varying ensemble
depth and number of estimators; we treat it as a directional observation
consistent with the data.  Regardless of mechanism, the lower XGBoost ESI
(e.g.\ $0.111$ vs.\ $0.287$ on Phishing) reinforces that robustness and
explanation stability are distinct axes that must be measured separately.

\subsection{Square Attack Adversarial Training for Tree Ensembles}
\label{sec:advtrain-trees}

Adversarial training for tree ensembles cannot follow the gradient-based recipe
used for MLPs, because Random Forests and XGBoost are non-differentiable and
do not admit FGSM or PGD augmentation.  As a black-box analogue we use Square
Attack data augmentation: generate perturbed training examples using Square
Attack at $\varepsilon_{\max}=0.3$, augment the original training set with
these adversarial copies ($25\%$ augmentation ratio, i.e.\ adding adversarial
examples equal to 25\% of the original training set), and retrain the model
on the combined set.  This mirrors the structural design of the MLP adversarial
training in \cref{sec:base}, though with a larger augmentation $\varepsilon$
($0.3$ here vs.\ $0.05$ for MLP FGSM) reflecting that the tree evaluation
operates at the full $\varepsilon_{\max}$ budget.

Table~\ref{tab:advtrain} reports the before/after comparison on Phishing and
UNSW-NB15 across all three attacks.  \textbf{XGBoost benefits most on
Phishing}: Square Attack RI rises from $0.372$ to $0.755$ ($+0.383$),
nearly closing the gap with RF ($0.769$).  Clean accuracy cost is negligible
($-0.001$).  XGBoost's piecewise-constant leaf structure makes it highly
susceptible to score-based random search; training on adversarial examples
forces a smoother partitioning.  As a side effect, ZOO degeneracy is partially
broken ($0.977 \to 0.928$): the retrained model can be fooled slightly under
ZOO, reflecting the smoother gradient surface.

On \textbf{UNSW-NB15}, the baseline RF collapses to near-zero accuracy at
$\varepsilon \geq 0.167$ (Square RI\,=\,0.184).  Adversarial augmentation
\emph{prevents this collapse}: accuracy remains $0.32$ at $\varepsilon=0.167$
and $0.07$ at $\varepsilon=0.3$, raising Square RI from $0.184$ to $0.340$
($+0.156$).  For XGBoost on UNSW-NB15, the improvement is more modest
($+0.075$ on Square), but cross-attack transfer is observed: ZOO RI improves
by $+0.045$ and HSJ RI by $+0.081$, suggesting genuine boundary hardening
rather than attack-specific tuning.

\textbf{Phishing RF is the exception}: Square RI decreases by $0.049$
($0.769 \to 0.720$) while ZOO and HSJ also decline marginally.  The RF already
achieves moderate Square robustness on Phishing; augmenting at the full
$\varepsilon_{\max}=0.3$ budget is too aggressive and shifts the boundary in
ways that hurt moderate-$\varepsilon$ attack performance.  For already-robust models, the augmentation $\varepsilon$ should be tuned
below $\varepsilon_{\max}$ ---
an analogue of the FGSM step-size sensitivity observed for MLP PGD.

\begin{table}[t]
\caption{Before/after RI comparison for Square Attack adversarial training
($25\%$ augmentation at $\varepsilon_{\max}=0.3$).
$\Delta$ is the change in RI for the adversarially trained model relative to
baseline.  $+$ indicates improvement; $-$ indicates degradation.}
\label{tab:advtrain}
\centering
\small
\begin{tabular}{llcccccccc}
\toprule
& & \multicolumn{4}{c}{Baseline RI} & \multicolumn{4}{c}{$\Delta$ (adv-trained $-$ baseline)} \\
\cmidrule(lr){3-6}\cmidrule(lr){7-10}
Dataset & Model & Clean & ZOO & Square & HSJ & $\Delta$Clean & $\Delta$ZOO & $\Delta$Sq & $\Delta$HSJ \\
\midrule
Phishing   & RF  & 0.982 & 0.901 & 0.769 & 0.920 & $-$0.004 & $-$0.052 & $-$0.049 & $-$0.024 \\
Phishing   & XGB & 0.987 & 0.977 & 0.372 & 0.876 & $-$0.001 & $-$0.049 & \textbf{$+$0.383} & $-$0.008 \\
UNSW-NB15  & RF  & 0.964 & 0.809 & 0.184 & 0.844 & $-$0.007 & $+$0.006 & \textbf{$+$0.156} & $\pm$0.000 \\
UNSW-NB15  & XGB & 0.974 & 0.886 & 0.299 & 0.806 & $-$0.003 & $+$0.045 & $+$0.075 & $+$0.081 \\
\bottomrule
\end{tabular}
\end{table}

\begin{figure}[htb]
\centering
\includegraphics[width=\textwidth,
  alt={Four panels arranged in a two-by-two grid. Top row: Phishing URL dataset, left panel Random Forest, right panel XGBoost. Bottom row: UNSW-NB15 dataset, left panel Random Forest, right panel XGBoost. Each panel shows two accuracy-versus-epsilon curves: a dashed blue line for the baseline model and a solid orange line for the adversarially trained model. The XGBoost Phishing panel shows the most dramatic shift, with the adversarially trained curve rising from near-random to moderate accuracy across the perturbation range. The UNSW panels show both models avoiding the near-zero accuracy collapse that baseline models exhibit at high epsilon.}
]{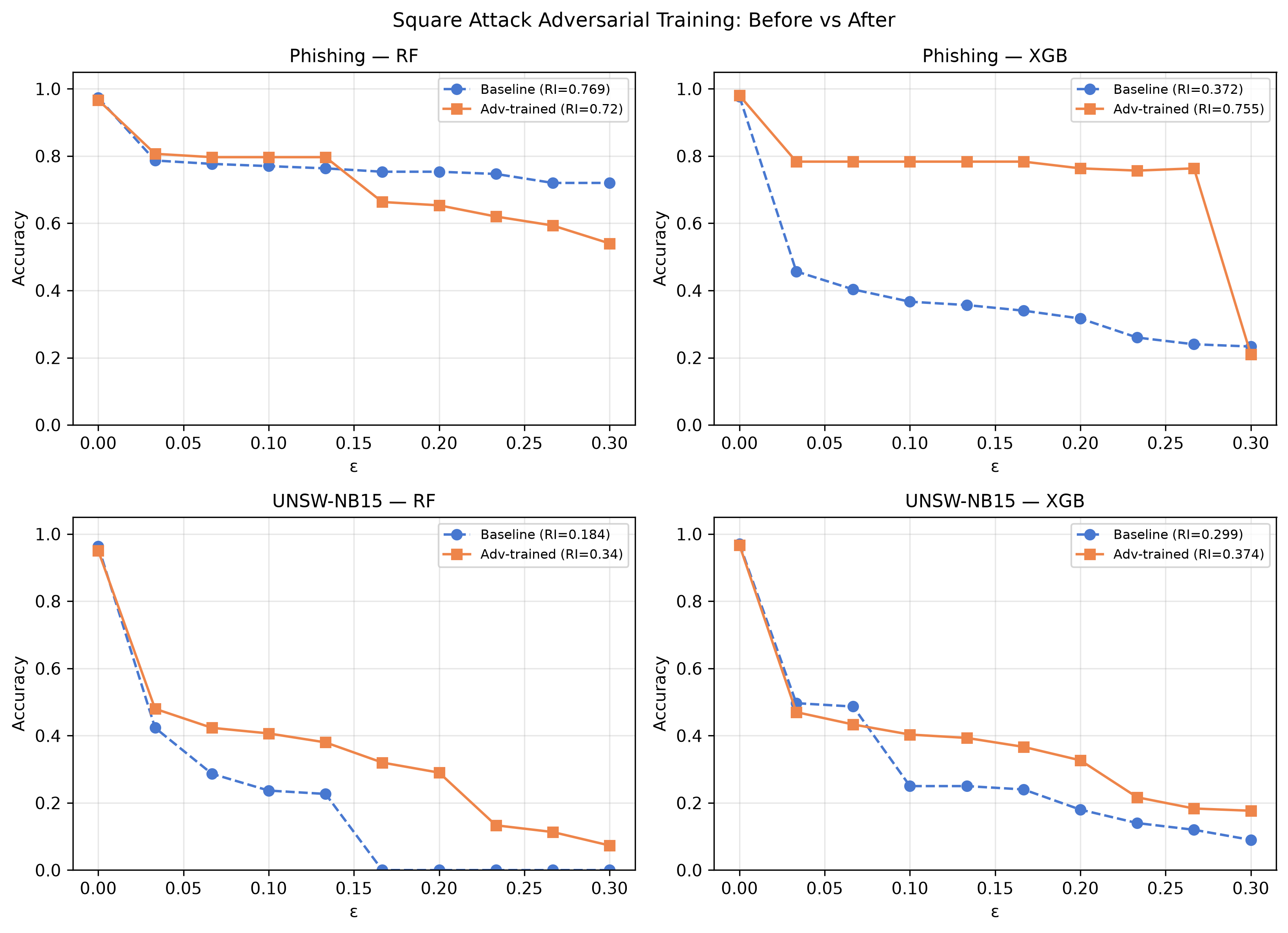}
\caption{Square Attack accuracy curves before (baseline, dashed blue) and after
(adv-trained, solid orange) augmentation on Phishing and UNSW-NB15.
Largest gains: XGBoost Phishing $+0.383$ RI, RF UNSW-NB15 $+0.156$ RI.}
\label{fig:advtrain-trees}
\end{figure}

\Cref{fig:advtrain-trees} plots the Square Attack accuracy curves before and
after augmentation on both datasets.  The XGBoost Phishing curve shifts
markedly upward.  For UNSW-NB15, both models' curves no longer drop to zero
at moderate $\varepsilon$.  The adversarially trained XGBoost on Phishing
achieves Square RI parity with the baseline RF ($0.755$ vs.\ $0.769$),
demonstrating that the large ZOO-Square gap can be substantially narrowed
through targeted augmentation.

\section{NF-ToN-IoT: Third Dataset Evaluation}
\label{sec:toniot}

Table~\ref{tab:master} includes results for NF-ToN-IoT.  Both RF and XGBoost
achieve 0.998 clean accuracy on this dataset, consistent with its highly separable
NetFlow feature structure.  Near-perfect clean accuracy on a sampled subset warrants caution: it could reflect genuine class separability in
NetFlow~v2 features, or it could indicate overfitting to the 40{,}000-sample
stratified subset or feature leakage from NetFlow-derived fields that encode
near-ground-truth labels.  Absolute RI values on this dataset should therefore
be treated with care; however, the \emph{relative} comparisons across attack
methods (e.g.\ ZOO degeneracy on XGBoost, Square Attack vs.\ HopSkipJump gap)
are less sensitive to clean-accuracy inflation and remain informative.
We flag this as a direction for further dataset validation.

A notable departure from Phishing and UNSW-NB15 is that ZOO RI for RF drops to
0.679 on NF-ToN-IoT, significantly lower than on the other datasets (0.915,
0.949).  One plausible explanation is the lower feature dimensionality of NF-ToN-IoT
(10 features vs.\ 48 for Phishing and 43 for UNSW-NB15): with fewer features,
ZOO's finite-difference gradient estimates sample a smaller coordinate space and
may be more accurate, reducing the degeneracy.  However, the three datasets also
differ in domain, feature-type distribution, and class separability, so
dimensionality cannot be cleanly isolated as the cause without a controlled
ablation that varies feature count within a single dataset.  We treat this as a
plausible hypothesis consistent with the data rather than a confirmed finding.

XGBoost ZOO RI remains high at 0.973, confirming that the piecewise-constant
prediction surface of boosted trees dominates regardless of dimensionality.
Square Attack degrades both models substantially (RI\,$\approx$\,0.46--0.48,
corresponding to terminal accuracy near the class-prior baseline),
consistent with prior datasets.

HopSkipJump achieves RI of 0.517~(RF) and 0.516~(XGB) on NF-ToN-IoT, converging
much closer to Square Attack than on the higher-dimensional Phishing and UNSW-NB15
datasets.  This convergence supports the interpretation that lower feature
dimensionality makes boundary search more tractable: with only 10 features,
HSJ's binary-search-and-gradient-estimate loop explores a substantially smaller
volume, approaching the effectiveness of Square Attack's random search.

ESI for RF (0.214) is between Phishing (0.287) and UNSW-NB15 (0.142);
XGBoost ESI (0.159) follows the same cross-dataset pattern.  However, the
five-seed variance study (\Cref{tab:variance}) shows the RF$-$XGB ESI margin
on NF-ToN-IoT is $0.025 \pm 0.082$, which is within sampling noise; this
ordering should be treated as inconclusive on this dataset.

\begin{figure}[t]
  \centering
  \includegraphics[width=0.85\textwidth,
    alt={Line plots showing adversarial robustness curves for Random Forest and XGBoost on the NF-ToN-IoT dataset under ZOO, Square Attack, and HopSkipJump attacks. Random Forest ZOO accuracy degrades more steeply than on other datasets (Robustness Index 0.679) due to the lower 10-feature dimensionality improving ZOO gradient estimates. XGBoost ZOO remains near-perfect. Square Attack degrades both models to near class-prior accuracy.}
  ]{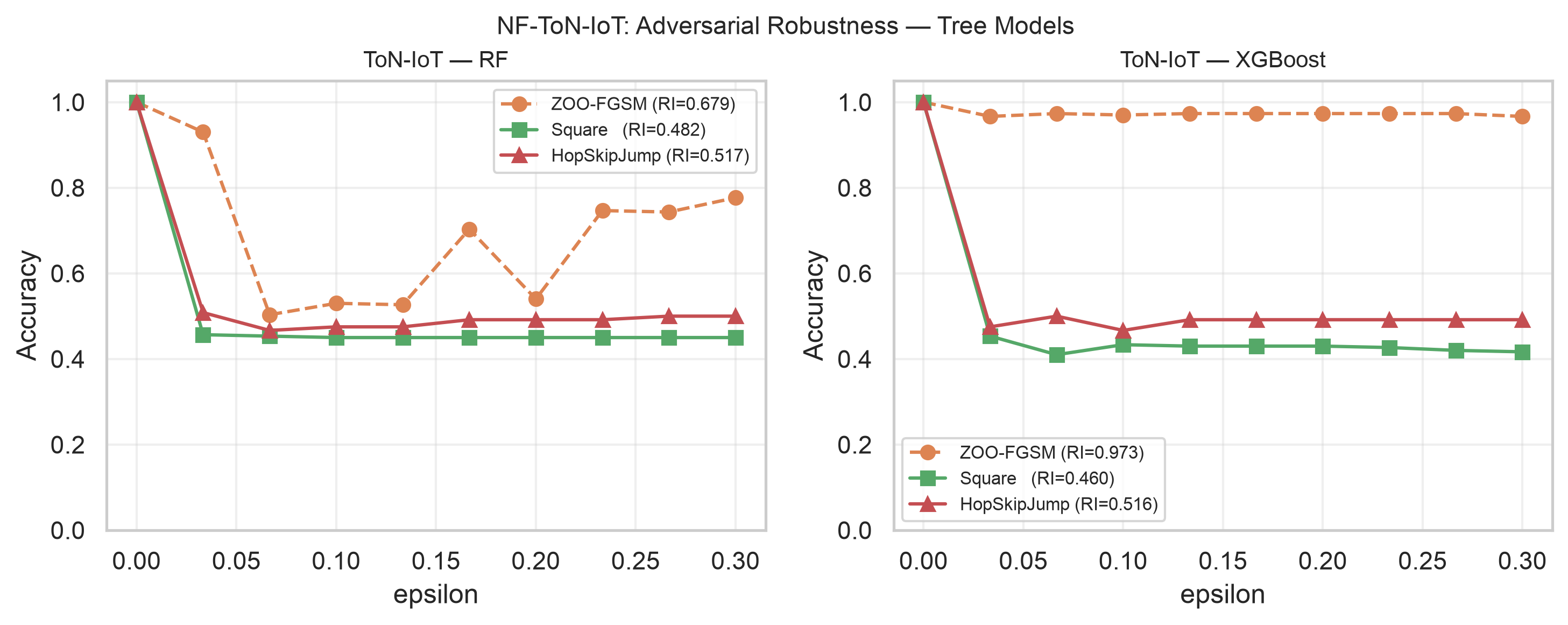}
  \caption{Adversarial robustness curves for RF and XGBoost on NF-ToN-IoT.
  ZOO is partially effective against RF (RI\,=\,0.679).  HopSkipJump converges
  close to Square Attack on both models (gap $\leq$\,0.06).}
  \label{fig:toniot}
\end{figure}

\Cref{fig:toniot} shows the robustness curves for both models on NF-ToN-IoT.

\section{HIKARI-2021: Cross-Domain Validation}
\label{sec:hikari}

HIKARI-2021~\cite{ferrag2022hikari} differs structurally from the three previous
datasets: it captures TLS/HTTPS encrypted network flows, meaning packet payload is
not directly observable, and its 83 retained features are dominated by connection
timing, byte/packet statistics, and TLS handshake characteristics rather than
URL-structure or categorical NetFlow fields.  It therefore represents a third
distinct security domain (encrypted-traffic detection) and tests whether the
findings from Phishing and UNSW-NB15 generalise beyond those specific feature
spaces.  As with NF-ToN-IoT, both RF and XGBoost achieve very high clean accuracy
(0.998 and 0.999); the same caution about relative comparisons over absolute RI
values applies here.

\paragraph{ZOO degeneracy persists.}
XGBoost ZOO RI is 0.997---the characteristic flat accuracy curve at clean
accuracy across all $\varepsilon$---confirming that piecewise-constant tree leaves
produce near-zero finite-difference gradients regardless of feature space.  The
ZOO-versus-Square-Attack gap for XGBoost is $0.997 - 0.438 = 0.559$, comparable
to the 0.62 gap on Phishing and consistent across all four datasets.

\paragraph{Square Attack reveals vulnerability.}
Under Square Attack, XGBoost RI drops to 0.438---meaning the model's accuracy
averaged across the $\varepsilon$ grid falls below $0.5$, indicating near- or
sub-random classification across most evaluated perturbation levels.
RF RI under Square Attack is 0.492, also near-random.  This is the most extreme
collapse we observe across all four datasets, suggesting that the encrypted-traffic
feature space provides a particularly dense adversarial neighbourhood under
$\ell_\infty$ perturbations.

\paragraph{HopSkipJump--Square gap on HIKARI.}
XGBoost HSJ RI is 0.894 versus Square Attack RI of 0.438---a gap of 0.456, larger
than on any other dataset.  With 83 features, HIKARI is high-dimensional, and
HSJ's boundary traversal must search a large volume; the result is consistent with
the trend established across previous datasets (larger gaps at higher
dimensionality).  RF shows a smaller HSJ--Square gap ($0.619 - 0.492 = 0.127$),
consistent with RF's smoother decision boundary making boundary search
comparatively more effective.

\paragraph{ESI ordering holds.}
RF ESI (0.140) exceeds XGBoost ESI (0.056) on HIKARI, extending the RF\,>\,XGB
ordering to a fourth dataset with a qualitatively different feature space.
XGBoost ESI of 0.056 is the lowest value observed across all datasets, tied with
UNSW-NB15 (0.063), reinforcing that XGBoost's additive boosting structure
produces higher attribution drift than Random Forest averaging irrespective of
the traffic domain.

\begin{figure}[t]
  \centering
  \includegraphics[width=0.85\textwidth,
    alt={Line plots showing adversarial robustness curves for Random Forest and XGBoost on the HIKARI-2021 dataset under ZOO, Square Attack, and HopSkipJump. XGBoost ZOO accuracy remains flat at near-clean accuracy (RI 0.997), demonstrating ZOO degeneracy. Square Attack reduces XGBoost RI to 0.438, meaning accuracy averaged across the perturbation grid falls below 0.5. Random Forest robustness curves degrade more uniformly across attack methods. HopSkipJump achieves a very high RI for XGBoost (0.894) despite the same dataset, reflecting the difficulty of boundary search in the 83-dimensional feature space.}
  ]{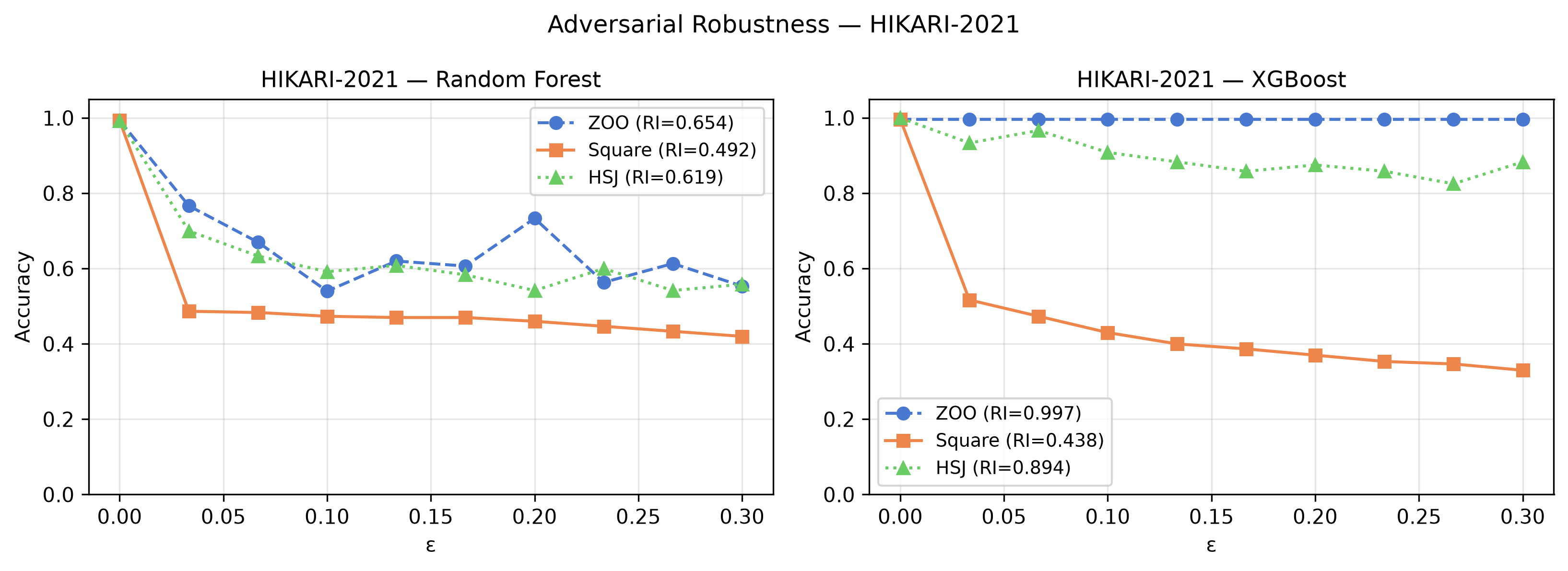}
  \caption{Adversarial robustness curves for RF and XGBoost on HIKARI-2021.
  XGB ZOO RI\,=\,0.997 (degeneracy); XGB Square RI\,=\,0.438.
  HSJ--Square gap for XGBoost (0.456) is the largest across all four datasets.}
  \label{fig:hikari}
\end{figure}

\Cref{fig:hikari} shows the robustness curves for both models on HIKARI-2021.

\section{PGD Step-Size Ablation}
\label{sec:pgd}

Our conference paper~\cite{khawarey2026empirical} observed that PGD produces
higher RI than FGSM---i.e., PGD is a \emph{weaker} attack---on both datasets.
Reviewer feedback identified this as counterintuitive, since PGD is an iterative
refinement of FGSM that should find stronger adversarial examples.

\begin{figure}[t]
  \centering
  \includegraphics[width=\textwidth,
    alt={Two line plots showing MLP accuracy versus perturbation magnitude epsilon for FGSM and PGD at five step sizes (alpha equals 0.005, 0.01, 0.02, 0.05, 0.10) with 10 steps fixed. Left: Phishing URL dataset; right: UNSW-NB15 dataset. PGD with small alpha (0.005, 0.01) is weaker than FGSM (higher accuracy under attack). As alpha increases to 0.05 and 0.10, PGD matches and surpasses FGSM effectiveness, confirming the step-size sensitivity artifact on z-score normalised tabular data.}
  ]{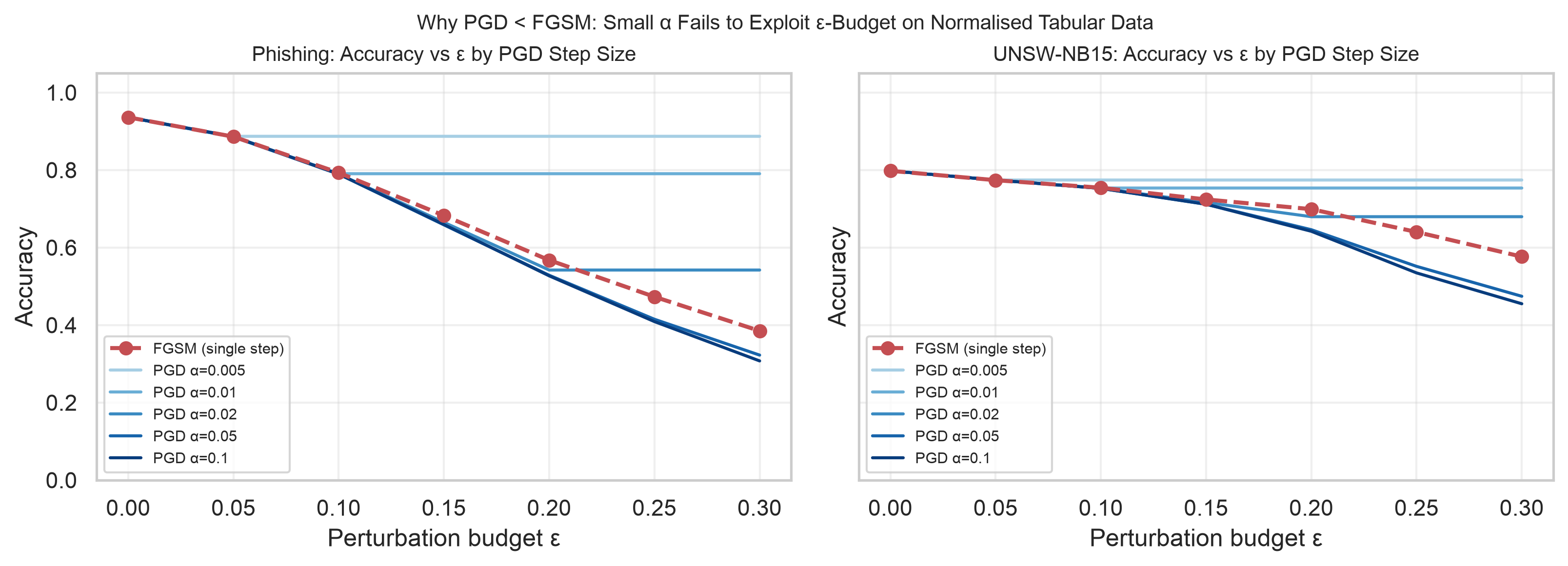}
  \caption{Accuracy vs.\ $\varepsilon$ for FGSM and PGD at five step sizes on
  Phishing (left) and UNSW-NB15 (right).  PGD with small $\alpha$ is a weaker
  attack than FGSM; the gap closes at $\alpha \geq 0.05$.}
  \label{fig:pgd_ablation}
\end{figure}

We resolve this by an ablation over five step sizes
$\alpha \in \{0.005,\allowbreak\,0.01,\allowbreak\,0.02,\allowbreak\,0.05,\allowbreak\,0.10\}$
with 10 fixed steps, shown in \Cref{tab:pgd_ablation} and \Cref{fig:pgd_ablation}.

\begin{table}[t]
\caption{PGD step-size ablation: Robustness Index (RI) on Phishing and UNSW-NB15
for MLP.  Higher RI indicates a weaker attack.  FGSM baseline shown for
reference.  Values differ from \Cref{tab:master} because the ablation uses
the full MLP test sets (no attack-budget subset sampling) and a fixed
$\varepsilon_{\max}=0.30$ grid, while \Cref{tab:master} evaluates on a
stratified 300-sample subset; both use the same trained model and seed.}
\label{tab:pgd_ablation}
\centering
\begin{tabular}{llcc}
\toprule
Attack & $\alpha$ & RI (Phishing) & RI (UNSW-NB15) \\
\midrule
FGSM        & ---   & 0.677 & 0.713 \\
\midrule
\multirow{5}{*}{PGD (10 steps)}
            & 0.005 & 0.891 & 0.776 \\
            & 0.010 & 0.819 & 0.761 \\
            & 0.020 & 0.694 & 0.724 \\
            & 0.050 & 0.651 & 0.679 \\
            & 0.100 & 0.649 & 0.674 \\
\bottomrule
\end{tabular}
\end{table}

The ablation reveals a monotonic decrease in RI as $\alpha$ increases: PGD
becomes progressively stronger as the step size grows, converging toward and
eventually surpassing FGSM at $\alpha \geq 0.05$.  The conventional choice of
$\alpha = \varepsilon_{\max}/\text{steps} = 0.01$ is the weakest PGD
configuration tested.

The mechanism is as follows.  On z-score normalised tabular data, each PGD step
of size $\alpha=0.01$ is projected back into the $L_\infty$ ball around
$\mathbf{x}_0$.  When $\varepsilon$ is small (e.g.\ $\varepsilon = 0.033$),
$10 \times 0.01 = 0.10 > \varepsilon$, so the cumulative displacement from 10
steps would overshoot the ball boundary and be clipped at each projection---the
iterate never reaches the boundary the single FGSM step jumps to directly.  At
larger $\varepsilon$ (e.g.\ $\varepsilon = 0.30$, $10 \times 0.01 = \varepsilon$),
each step fits within the ball and PGD can accumulate its full budget, recovering
FGSM-level strength.  FGSM's single full-budget step reaches the $L_\infty$
boundary directly at all $\varepsilon$ values without iterative clipping.  This phenomenon is specific to tabular
data with z-score normalisation, where feature scales are uniform and small step
sizes are systematically ineffective.  We recommend adaptive step sizes
($\alpha = \varepsilon / \text{steps}$ per $\varepsilon$ value) for future
studies on normalised tabular data.

\section{Discussion}
\label{sec:discussion}

\subsection{Attack Realism for Structured Security Features}

A practical concern for adversarial evaluation on tabular cybersecurity data is
whether $L_\infty$-bounded perturbations correspond to plausible adversarial
capabilities.  In phishing detection, features represent observable URL and page
structure; an attacker can realistically manipulate these (e.g.\ adding subdomains,
adjusting URL length) while preserving the phishing payload.  In network intrusion
detection, NetFlow features such as byte counts and packet rates can be shaped by
an attacker controlling the traffic source.  However, not all features are
independently actionable: some are deterministically derived from others (e.g.\
packet counts constrain byte counts), meaning unconstrained $L_\infty$ perturbations
may generate invalid feature combinations.

All five attacks in this study apply unconstrained $\ell_\infty$ perturbations
without enforcing domain-specific feature validity.  Realism is not a function
of attack sparsity or perturbation structure: Square Attack's random square-shaped
patches do not, by construction, preserve semantic constraints any more than FGSM
or ZOO do.  The lower RI values under Square Attack and HopSkipJump relative to
ZOO reflect differences in \emph{attack effectiveness}---principally the ZOO
degeneracy artefact---not differences in how closely the perturbations respect
the feature manifold.

A genuine realism argument for tabular security data requires distinguishing
attacker-controllable features from derived or protocol-constrained ones.  In
phishing detection most URL-structure features are directly under the attacker's
control (the attacker constructs the phishing page), which provides a reasonable
basis for unconstrained perturbations.  In network intrusion detection, however,
some NetFlow features are deterministic functions of others (packet counts
constrain byte-per-packet ratios), so any perturbation that violates these
constraints produces records no real flow could generate.  Making the attack protocol domain-realistic requires projecting perturbations
onto the manifold of valid network records---a direction formalised as the
\emph{problem-space} constraint by Pierazzi et al.~\cite{pierazzi2020intriguing}
and left for future work in our setting~\cite{apruzzese2022modeling}.

\subsection{ESI as a Complement to RI}

RI and ESI measure distinct properties that can decouple.  A model
may retain its prediction under perturbation while its attributions reorder
completely; conversely, stable attributions do not guarantee a stable prediction.
ESI therefore supplies information that RI alone cannot, capturing the
trustworthiness of a model's explanations under attack rather than the
correctness of its label.

This positioning distinguishes ESI from existing explanation-stability metrics.
Evaluation frameworks such as OpenXAI~\cite{agarwal2022openxai} formalise
relative input and output stability (RIS/ROS), which quantify the faithfulness
and local Lipschitz behaviour of explanations---how much an explanation changes
under random or local input noise.  Alvarez-Melis and
Jaakkola~\cite{alvarezmelis2018robust} similarly probe explanation sensitivity
to local perturbations.  ESI instead measures attribution drift under
\emph{adversarial} perturbations of growing magnitude, tying explanation
stability to the same threat model that defines RI.  Where existing metrics ask
whether explanations are locally smooth, ESI asks whether they survive an
adversary.

\begin{figure}[htb]
  \centering
  \includegraphics[width=0.85\textwidth,
    alt={Two line plots showing mean absolute SHAP attribution drift versus perturbation magnitude for Random Forest (left) and XGBoost (right) on the Phishing dataset. Each plot shows two curves: ZOO-based ESI and Square Attack-based ESI. For both models, drift under Square Attack is higher than under ZOO, confirming that ZOO underestimates perturbation strength. Across both attacks, Random Forest ESI remains higher than XGBoost ESI, confirming the ordering is architectural rather than a ZOO artefact.}
  ]{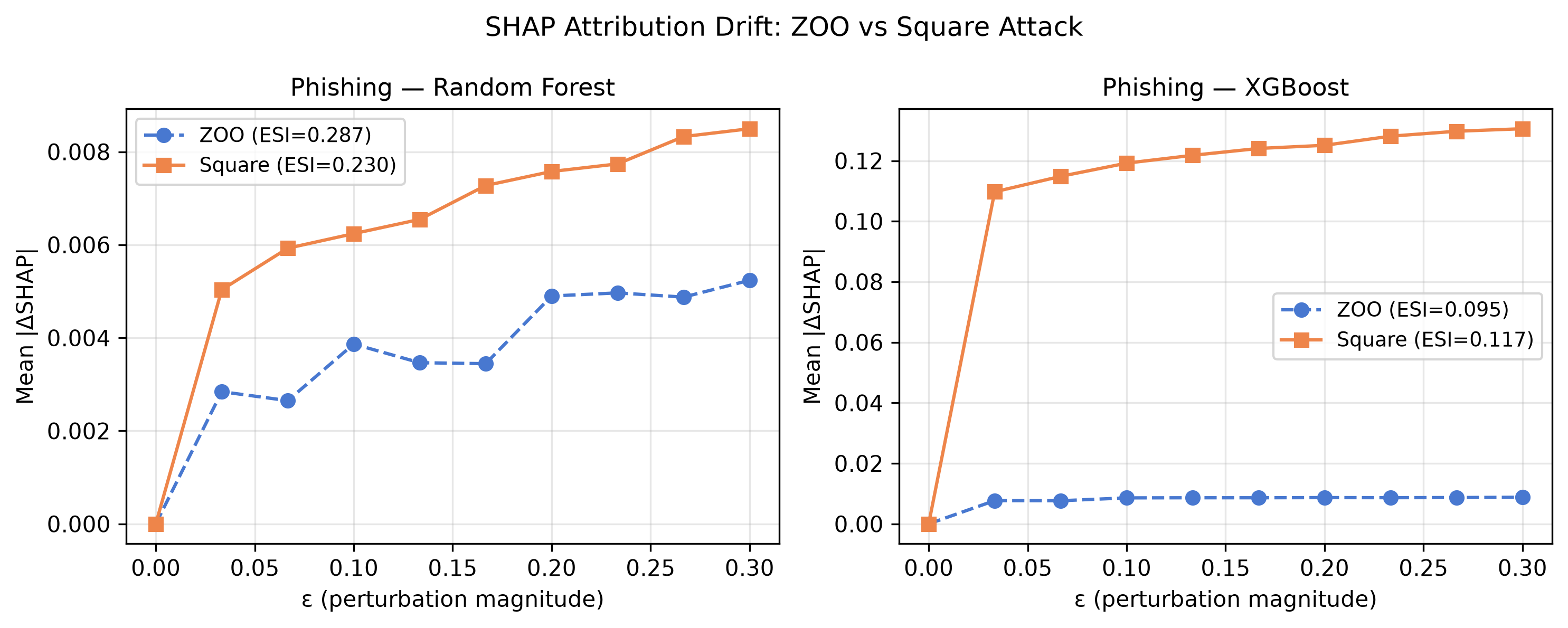}
  \caption{SHAP attribution drift under ZOO and Square Attack on Phishing.
  RF ESI\,=\,0.287 (ZOO) / 0.230 (Square); XGB ESI\,=\,0.111 (ZOO) / 0.117
  (Square).  RF\,>\,XGB under both attacks.}
  \label{fig:esi_square}
\end{figure}

A natural concern is whether the RF\,>\,XGBoost ESI ordering is genuine or an
artefact of ZOO degeneracy: since ZOO barely perturbs XGBoost predictions, the
lower XGBoost ESI might reflect that ZOO perturbs the two models to structurally
different degrees rather than a true architectural difference in explanation
fragility.  To test this, we computed ESI under Square Attack---which actually
moves XGBoost predictions---on the Phishing dataset.  The ordering holds: RF
ESI\,=\,0.230 versus XGBoost ESI\,=\,0.117 under Square Attack, compared to
$0.287$ vs.\ $0.111$ under ZOO.  XGBoost explanation fragility persists even
when the generating attack is not degenerate, confirming that the gap reflects an
architectural property, not a ZOO artefact.
\Cref{fig:esi_square} shows the attribution drift curves under both attacks.

\begin{figure}[htb]
  \centering
  \includegraphics[width=0.85\textwidth,
    alt={Two line plots showing mean absolute SHAP attribution drift versus perturbation magnitude for Random Forest (left) and XGBoost (right) on the UNSW-NB15 dataset. Each plot shows two curves: ZOO-based ESI and Square Attack-based ESI. For both models, drift under Square Attack is higher than under ZOO. Random Forest ESI under Square Attack (0.184) exceeds XGBoost ESI (0.155), confirming that the RF greater than XGB ordering is not specific to the Phishing dataset.}
  ]{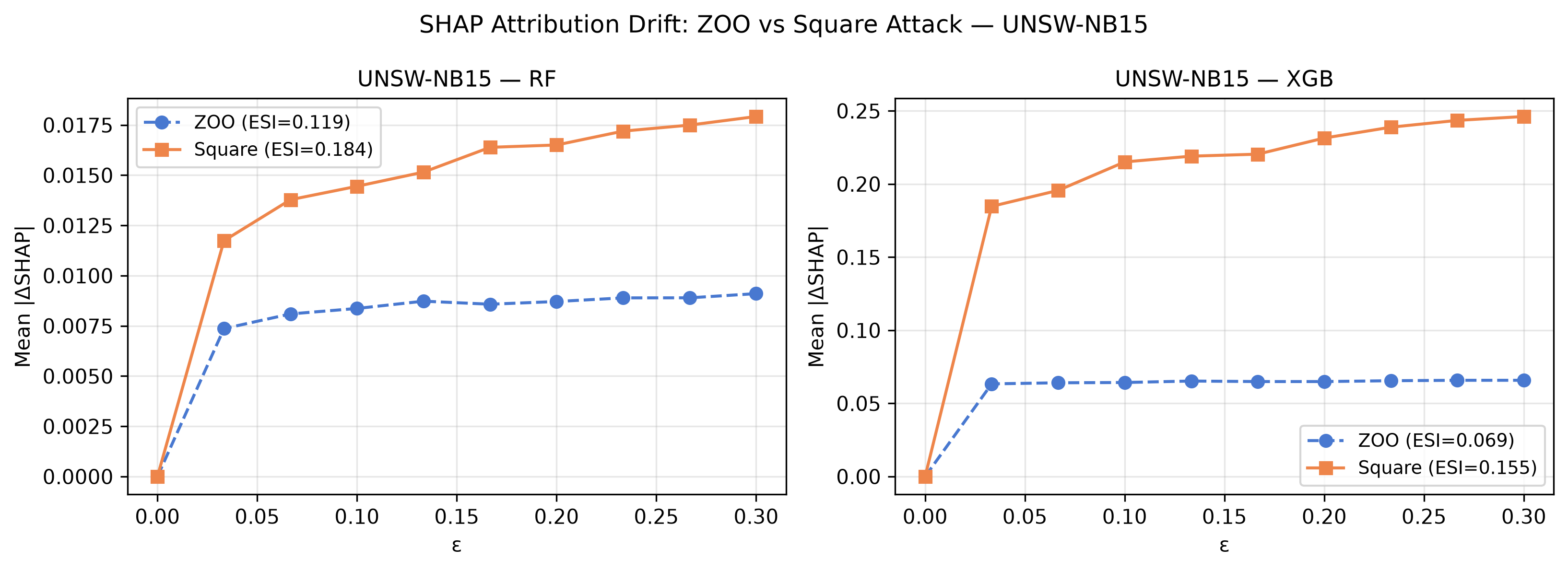}
  \caption{SHAP attribution drift under ZOO and Square Attack on UNSW-NB15.
  RF Square ESI\,=\,0.184; XGB Square ESI\,=\,0.155.}
  \label{fig:esi_square_unsw}
\end{figure}

The cross-dataset behaviour of ESI supports its use as a model-selection
criterion.  Across all four datasets, XGBoost records a lower ZOO ESI than the
Random Forest ($0.111$ vs.\ $0.287$ on Phishing, $0.068$ vs.\ $0.142$ on
UNSW-NB15, $0.159$ vs.\ $0.214$ on NF-ToN-IoT, and $0.056$ vs.\ $0.140$ on
HIKARI-2021).  Under Square Attack, the RF\,>\,XGB ordering holds on UNSW-NB15:
RF\,ESI\,=\,0.184 versus XGB\,ESI\,=\,0.155, corroborating the architectural
interpretation on a second, structurally distinct dataset
(\Cref{fig:esi_square_unsw}).  On NF-ToN-IoT the margin is negligible
(RF\,$\approx$\,XGB\,$\approx$\,0.105 in a single run).  On HIKARI-2021 the
ordering reverses (RF\,ESI\,=\,0.093, XGB\,ESI\,=\,0.176)---a normalisation-scale
artefact: XGBoost's absolute mean SHAP drift on HIKARI ($0.063$--$0.101$ per
feature) is approximately ten times RF's ($0.008$--$0.010$), yet it grows
gradually, starting at 63\% of its maximum at $\varepsilon=0.033$ and reaching
100\% at $\varepsilon=0.30$, while RF's drift saturates at 86\% of its maximum
at the first $\varepsilon$ step.  The ESI normalisation by
$\bar{D}(\varepsilon_{\max})$ (Eq.~\ref{eq:esi}) rewards gradual drift curves
regardless of absolute scale, so XGBoost scores as \emph{more} stable despite
its far larger absolute attribution change.  Normalised-AUC metrics should
therefore be read alongside absolute drift magnitudes when model families
produce SHAP values of different scales.  In sum: the RF\,>\,XGB ordering holds
on all four datasets under ZOO; under Square Attack it is confirmed on Phishing
and UNSW-NB15, inconclusive on NF-ToN-IoT, and reversed on HIKARI-2021 due to
the normalisation artefact above.

Low ESI is operationally hazardous even when predictions appear stable---whether
from genuine robustness or from attack degeneracy.  A SOC analyst relying on
SHAP attributions from such a model may be steered toward the wrong features
and the wrong remediation, even as the predicted label remains correct.
Reporting RI and ESI jointly, and under multiple attack methods to avoid
degeneracy artefacts, lets practitioners identify models that are robust in
label but fragile in rationale.  Future work will extend ESI beyond SHAP to
LIME~\cite{ribeiro2016lime} attributions, enabling explanation-stability
comparisons across interpretation methods as well as across models.

\subsection{Attack-Method Selection Guidance}
\label{sec:attack_guidance}

The results across four datasets and two tree model families reveal a consistent
ranking of attack strength that can be explained by two independent axes:
\emph{gradient dependence} and \emph{query efficiency in high-dimensional spaces}.

\paragraph{Gradient dependence.}
ZOO's effectiveness is determined by the quality of its finite-difference
gradient estimates, which in turn depends on model smoothness and feature
dimensionality.  Against XGBoost, whose piecewise-constant prediction surface
yields near-zero finite differences inside tree leaves, ZOO is essentially
degenerate at all dimensionalities (RI\,=\,0.97--0.99 across all four datasets).
Against Random Forest, whose ensemble averaging produces smoother decision
boundaries, ZOO is partially effective but strongly dimensionality-dependent:
RI drops from 0.949 (UNSW-NB15, 43 features) to 0.679 (NF-ToN-IoT, 10 features).
Square Attack and HopSkipJump require no gradient information and are unaffected
by this axis entirely.

\paragraph{Query efficiency.}
Given equivalent query budgets, Square Attack's random $\ell_\infty$ search
outperforms HopSkipJump's boundary traversal in high-dimen\-sional feature spaces.
The RI gap is largest for XGBoost on Phishing (Square Attack 0.36 vs.\ HSJ 0.65,
a gap of 0.29) and closes at low dimensionality: on NF-ToN-IoT (10 features)
both methods converge to RI\,$\approx$\,0.48--0.52 for both models.  This is
consistent with boundary search being more adversely affected by dimensionality
than random search.

\paragraph{Practical recommendations.}
These two axes suggest the following guidelines for adversarial evaluation of
cybersecurity classifiers:
\begin{itemize}
  \item \textbf{Tree ensembles (XGBoost, RF):}  Square Attack should be the
    primary benchmark.  ZOO alone is insufficient and can overestimate robustness
    by up to 0.62 RI for XGBoost.  HopSkipJump provides useful complementary
    evidence that the attack is decision-boundary-grounded, but requires a
    larger query budget and may be weaker in high-dimensional feature spaces.
  \item \textbf{MLP (white-box):}  FGSM or PGD with adaptive step size
    ($\alpha = \varepsilon/\text{steps}$ per $\varepsilon$ value) should be
    preferred over default PGD ($\alpha = 0.01$), which underperforms FGSM on
    z-score normalised tabular data.
  \item \textbf{Robustness reporting:}  Report RI under the \emph{strongest
    available attack} as the primary robustness estimate.  Reporting ZOO RI
    alongside Square Attack RI is useful to quantify the degeneracy gap, but
    the headline figure should come from the stronger attack.
\end{itemize}

\section{Limitations}
\label{sec:limitations}

Several limitations bound the scope of our conclusions and motivate future work.

\paragraph{Feature validity.}
Our attacks apply $\ell_\infty$ perturbations independently to each feature,
which can violate the dependencies inherent in structured security data.  In
NetFlow records, for example, byte and packet counters are strongly correlated
and the ratio of \texttt{IN\_BYTES} to \texttt{IN\_PKTS} is bounded by physical
packet sizes, so an unconstrained perturbation may produce records that no real
flow could generate.  The reported robustness values are therefore conservative
upper bounds on realistic adversarial difficulty; future work should incorporate
domain-constraint projections that restrict perturbations to the manifold of
valid traffic.

\paragraph{Dataset scope.}
We evaluate four datasets drawn from three domains---web security (Phishing URLs),
network intrusion detection (UNSW-NB15, NF-ToN-IoT), and encrypted-traffic
detection (HIKARI-2021).  Other important security tasks, notably static and
dynamic malware classification and host-based intrusion detection, are not
covered, and their feature semantics may interact differently with both attacks
and explanation drift.

\paragraph{Computational constraints.}
All experiments run on CPU, which limits attack strength.  We cap HopSkipJump
at $120$ samples, against $300$ for the other attacks, and restrict the number of
ZOO iterations to keep evaluation tractable.  A GPU-based evaluation would permit
larger query budgets and more iterations, likely yielding lower RI values and thus
exposing stronger vulnerabilities than those reported here.  A query-budget
ablation (\cref{sec:trees}) directly tests whether ZOO degeneracy on XGBoost is
a budget artefact: varying the coordinate-sampling fraction from 12\% to 100\%
and the number of iterative refinement steps from 20 to 200 leaves XGBoost RI
flat on Phishing ($\Delta = 0.000$), establishing degeneracy as intrinsic to
the piecewise-constant prediction surface rather than a function of the iteration
limit used here.  GPU-scale budgets are unlikely to change this conclusion on
Phishing; on UNSW-NB15, where partial sensitivity was observed, higher budgets
may narrow the ZOO--Square gap further.

\paragraph{Variance and statistical significance.}
We ran each experiment across five random seeds that jointly control the
train/test split, model initialisation, and evaluation subset selection.
\Cref{tab:variance} reports mean\,$\pm$\,std for the two key comparisons.

\begin{table}[t]
\caption{Variance across 5 random seeds.  ``XGB gap'' is XGB ZOO\,RI $-$
XGB Square\,RI; ``RF$-$XGB ESI'' is the ESI margin between model families.
$k/5$ counts seeds where RF\,>\,XGB ESI holds.}
\label{tab:variance}
\centering
\small
\begin{tabular}{lccc}
\toprule
Dataset & XGB gap (ZOO$-$Sq RI) & RF$-$XGB ESI & RF$>$XGB ($k/5$) \\
\midrule
NF-ToN-IoT  & $0.541 \pm 0.023$ & $0.025 \pm 0.082$ & 3/5 \\
Phishing    & $0.575 \pm 0.029$ & $0.125 \pm 0.087$ & 4/5 \\
HIKARI-2021 & $0.537 \pm 0.017$ & $0.064 \pm 0.068$ & 4/5 \\
UNSW-NB15   & $0.534 \pm 0.035$ & $0.067 \pm 0.015$ & 5/5 \\
\bottomrule
\end{tabular}
\end{table}

The ZOO degeneracy gap is highly reproducible: std\,$\leq$\,0.035 across all
four datasets, confirming this is a structural property rather than a lucky
point estimate.  ESI is more variable: RF ESI std reaches 0.084--0.101 on
low-dimensional datasets (NF-ToN-IoT, Phishing), reflecting sensitivity of the
normalised-AUC formula to which 256 test examples are sampled.  The RF\,>\,XGB
ESI ordering holds in all five seeds on UNSW-NB15 (margin $0.067 \pm 0.015$,
the smallest relative std), in four of five seeds on Phishing and HIKARI-2021,
and in three of five on NF-ToN-IoT.  The NF-ToN-IoT margin ($0.025 \pm 0.082$)
is within noise and should be treated as inconclusive; the ordering on the other
three datasets is directionally consistent but confidence intervals overlap on
Phishing and HIKARI, so the architectural claim is best supported by the
qualitative pattern across datasets rather than any single point estimate.

\paragraph{Tree ensemble defences.}
\Cref{sec:advtrain-trees} evaluates Square Attack data augmentation as a
black-box adversarial training strategy for tree ensembles.  The largest gains
are observed where the baseline Square RI is lowest: XGBoost on Phishing
improves by $+0.383$, and RF on UNSW-NB15 by $+0.156$.  A remaining limitation
is that augmentation at the full $\varepsilon_{\max}=0.3$ budget can degrade
models that already have moderate robustness (RF on Phishing, $-0.049$), suggesting
the augmentation $\varepsilon$ should be cross-validated rather than fixed at
$\varepsilon_{\max}$.  Defence strategies beyond augmentation---such as
feature-validity constraint projections~\cite{apruzzese2022modeling}, ensemble
diversity maximisation, and adversarial purification via density estimators---remain
active research directions not covered here.

\paragraph{ESI approximation.}
For tree models we compute ESI using ZOO-perturbed inputs, which is convenient
because ZOO produces smoothly increasing perturbations at each $\varepsilon$
step, but it is not necessarily the attack that maximises attribution drift.
We have validated the ZOO-based ESI ordering under Square Attack on all four
datasets (see \cref{sec:discussion}): the RF\,>\,XGB ordering holds on Phishing
and UNSW-NB15, is tied on NF-ToN-IoT, and reverses on HIKARI-2021 due to a
normalisation-scale artefact.  A remaining limitation is that the validation
uses the same query budget as the RI evaluation; a drift-maximising attack
with a larger budget may rank model families differently.

\section{Conclusion}
\label{sec:conclusion}

We presented an extended empirical study of adversarial robustness and
explainability stability in tabular cybersecurity classifiers, expanding our
conference paper~\cite{khawarey2026empirical} along four dimensions: classifier
families (RF and XGBoost), attack methods (ZOO, Square Attack, HopSkipJump),
evaluation datasets (NF-ToN-IoT and HIKARI-2021), and metrics (ESI).

Three findings stand out.  First, ZOO produces degenerate gradient estimates
against XGBoost, inflating apparent robustness; Square Attack reveals genuine
vulnerability with a RI gap of up to 0.62 on the same model.  This gap
replicates across all four datasets (mean 0.534--0.575, std\,$\leq$\,0.035
across five seeds), confirming that attack-method selection is not model-agnostic
and that gradient-based black-box attacks are insufficient for evaluating tree
ensembles.  Second, RI and ESI can decouple: ZOO
perturbations that fail to fool XGBoost predictions (artefact RI\,$\approx$\,0.98)
still drive large attribution drift (ESI\,$\approx$\,0.06--0.16), demonstrating
that an attack too weak to cause misclassification can nonetheless destabilise
explanations.  RF achieves higher ESI than XGBoost under ZOO across all four datasets;
Square Attack recomputation confirms this ordering on Phishing
(RF\,=\,0.230, XGB\,=\,0.117) and UNSW-NB15 (RF\,=\,0.184, XGB\,=\,0.155),
corroborating the interpretation that the gap is architectural rather than a
ZOO artefact~(\Cref{fig:esi_square,fig:esi_square_unsw}); UNSW-NB15 provides
the strongest confirmation (5/5 seeds, margin $0.067 \pm 0.015$).
Third, PGD underperforms FGSM on z-score normalised tabular data when the
conventional step size $\alpha=0.01$ is used, an artifact explained by iterative
$L_\infty$ projection cancelling gradient accumulation.

Directions for future work include incorporating feature validity constraints into
the attack protocol, extending ESI to other XAI methods beyond SHAP (e.g.\ LIME),
and tuning the augmentation $\varepsilon$ budget for Square Attack adversarial
training---the current study uses $\varepsilon_{\max}=0.3$, but cross-validating
a smaller augmentation budget may improve gains for models with already-moderate
baseline robustness.

\begin{credits}
\subsubsection{\ackname}
The authors thank the anonymous reviewers of the ICAART 2026 conference version
for feedback that directly motivated the tree ensemble extension, the black-box
attack comparison, and the PGD step-size ablation presented in this work.
Experiment code and result tables are publicly available~\cite{khawarey2026code}.

\subsubsection{\discintname}
The authors have no competing interests to declare that are relevant to the
content of this article.
\end{credits}

\bibliographystyle{splncs04}
\bibliography{refs}

\end{document}